%% file: Main_text/main.tex
\documentclass[journal]{IEEEtran} 

\input{Main_text/Configuration_files/config}

\begin{document}

\input{Main_text/Sections/Title}

\input{Main_text/Sections/Abstract}

\input{Main_text/Sections/0_Introduction}
\input{Main_text/Sections/1_Background_TSP}
\input{Main_text/Sections/2_Domain_and_Edge_Signals}
\input{Main_text/Sections/3_Application}
\input{Main_text/Sections/4_Discussion}
\input{Main_text/Sections/5_Conclusions}
\input{Main_text/Acknowledgements}
\bibliographystyle{IEEEtran} 
\bibliography{bibliography.bib} 

\end{document}

%% file: Main_text/Configuration_files/config.tex
\usepackage{amsmath,amsfonts}
\usepackage{algorithmic}
\usepackage{algorithm}
\usepackage{array}
\usepackage[caption=false,font=normalsize,labelfont=sf,textfont=sf]{subfig}
\usepackage{textcomp}
\usepackage{stfloats}
\usepackage{url}
\usepackage{verbatim}
\usepackage{graphicx}
\usepackage{cite}
\usepackage{tcolorbox} 
\newtcolorbox{insightbox}[2][]{
    colback=cyan!2!white,     
    colframe=cyan!8!white,   
    fonttitle=\bfseries,       
    coltitle=black!70!black,   
    title=#2,                 
    #1
}

\usepackage{titlesec}

\usepackage{amsthm}
\newcommand{\vc}[1]{{\mathbf #1}}

\newcommand{\ma}[1]{{\mathbf #1}}

\newcommand{\transpose}{^{\mathrm{T}}}

\newcommand{\llow}{\mathbf{L}_{\downarrow}}
\newcommand{\lup}{\mathbf{L}_{\uparrow}}

\newcommand{\bk}{\mathbf{B}_k}
\newcommand{\bkt}{\mathbf{B}_k^{T}}
\newcommand{\bkpu}{\mathbf{B}_{k+1}}
\newcommand{\bkput}{\mathbf{B}_{k+1}^{T}}

\usepackage[acronym]{glossaries}
\newacronym{gso}{GSO}{graph shift operators}
\newacronym{gsp}{GSP}{graph signal processing}
\newacronym{dsp}{DSP}{digital signal processing}
\newacronym{tsp}{TSP}{topological signal processing}
\newacronym{tft}{TFT}{topological Fourier transform}
\newacronym{tso}{TSO}{topological shift operators}
\newacronym{tnn}{TNN}{Topological Neural Network}
\newacronym{mri}{MRI}{Magnetic Resonance Imaging}
\newacronym{fmri}{fMRI}{Functional Magnetic Resonance Imaging}
\newacronym{bold}{BOLD}{Blood-Oxygenation-Level-Dependent}
\newacronym{fc}{FC}{Functional Connectivity} 
\newacronym{sc}{SC}{Structural Connectivity}
\newacronym{ets}{eTS}{Edge Time Series}
\newacronym{hcp}{HCP}{Human Connectome Project}
\newacronym{mtd}{MTD}{Multiplication of Time Derivatives}
\newacronym{dwi}{DWI}{Diffusion Weighted Imaging}
\newacronym{leida}{LEiDA}{Leading Eigenvector Dynamics Analysis}
\newacronym{tr}{TR}{repetition time}
\newacronym{tda}{TDA}{Topological Data Analysis}
\newacronym{tdl}{TDL}{Topological Deep Learning}
\newacronym{tml}{TML}{Topological Machine Learning}

%% file: Main_text/Sections/Title.tex
\title{Topological Signal Processing: \\An Application-Oriented Tutorial}
\author{Flavia Petruso, 
        Maria Giulia Preti,~\IEEEmembership{Senior Member,~IEEE,},
        Dimitri Van De Ville,~\IEEEmembership{Fellow,~IEEE}.
\thanks{The authors are with the Neuro-X Institute, \'{E}cole Polytechnique F\'{e}d\'{e}rale de Lausanne (EPFL), Geneva, Switzerland, and the Department of Radiology and Medical Informatics, University of Geneva, Geneva, Switzerland. E-mail: flavia.petruso@epfl.ch}
\thanks{Manuscript received XXX; revised XXX.}
        \vspace*{-4ex}
}
\maketitle

%% file: Main_text/Sections/Abstract.tex
\begin{abstract}
    Many modern datasets are large and carry complex structural relationships. Graph-based methods have traditionally been used to represent networked data, modeling individual elements as nodes and pairwise interactions as edges. Furthermore, Graph Signal Processing (GSP) has been developed to analyze signals on graph nodes, such as temperature measurements (node signals) across different regions of a country represented as a graph. Topological Signal Processing (TSP) is an emerging field that generalizes GSP, enabling the analysis of signals defined not only on nodes but also on edges, triangles, and higher-dimensional network elements, modeled as simplicial complexes and related topological structures. This makes TSP naturally well-suited for studying higher-order interactions in complex systems by extending classical signal processing concepts, such as filtering and Fourier transforms, to the topological level. Despite its versatility, TSP remains challenging for many practitioners. Therefore, we present an accessible overview of TSP foundations while drawing connections with application-oriented settings. We focus on processing techniques based on the combinatorial Hodge Laplacian, which generalizes the graph Laplacian to simplicial complexes. In particular, we review key TSP concepts, relate them to real-world examples, and discuss how higher-order structures and signals can be derived from datasets. For instance, we introduce an edge-level signal capturing lagged interactions between nodal signals, and demonstrate its use in a case study on TSP-based analysis of brain imaging data, revealing nontrivial interactions between sets of brain regions. Overall, we aim to promote a broader adoption of TSP by bridging methodological developments with applications, fostering its use among a wide community of theoretical and applied researchers.
\end{abstract}

\begin{IEEEkeywords}
    Topological Signal Processing, Simplicial Complexes, Hodge Laplacian, Spectral Theory, Topological Signals, Edge Signals, Neuroimaging.
\end{IEEEkeywords}

%% file: Main_text/Sections/0_Introduction.tex
\section{Introduction and Motivations}
\label{sec:Introduction}
\IEEEPARstart{O}{ver} the past two decades, data availability has witnessed exponential growth representing complex structure residing on irregular domains such as spatial, social, or biological networks. This has led to the emergence of network science, a flurry of approaches that exploit structural and geometric properties of such networked data with a broad range of extensive applications \cite{barabasi2013network}. In this context, graphs have proven to be an extremely effective tool, as they naturally encode individual entities as nodes and their pairwise relationships as edges \cite{barabasi2013network, van2010graph, lewis2011network, newman2012communities}.

In many cases, however, data are not only structured as a network, but also measured on top of it. Examples include dynamic interactions among users in a social network \cite{singh2024social} or functional brain activity observed over networks of brain regions \cite{huang2018graph}. This perspective has fostered the development of \gls{gsp}, a branch of signal processing where a signal resides on the nodes of a graph and its processing is guided by the properties of the network \cite{ortega2018graph}. In practice, this is achieved by redefining the traditional concepts of neighborhood and adjacency in terms of the graph connectivity patterns, which are then represented algebraically using matrices such as the adjacency matrix and the graph Laplacian. By treating these matrices as graph-aware operators on signals, key signal processing operations such as filtering and Fourier transforms can be generalized to the graph setting \cite{leus2023graph}. This analysis enables the identification of properties such as smoothness or dominant patterns of variation that are consistent with the underlying domain structure. 

\subsection{Topological Signal Processing: Motivations, Fundamentals, and Recent Advances}
Overall, \gls{gsp} has driven major advances in analyzing signals on irregular data and networks, and has become a well-established yet still expanding research field with a vast range of applications \cite{ortega2018graph, leus2023graph}. Recent work has also pointed out the high expressive power of graph-based models, that can accommodate higher-order interactions through nonlinear functions of multiple graph elements \cite{peixoto2026graphs}. Such interactions naturally arise in a variety of applications, such road traffic, financial transactions, and biological or citation networks; see \cite{isufi2025topological,bick2023higher}, or \cite{battiston2020networks} for a list of examples. 

Nevertheless, classical \gls{gsp} is most naturally formulated for scalar signals defined on the nodes of a graph, while many relevant settings involve quantities associated with edges (e.g., flows on networks), which introduce an inherent notion of orientation absent at the nodal level, thereby adding an additional layer of modeling complexity. Furthermore, in several applications, the underlying domain may exhibit higher-order organization that is not easily or parsimoniously captured by graph-based models, while benefiting from an explicit representation of higher-order elements. These considerations have motivated the study of extensions of \gls{gsp} and alternative representations that provide complementary analytical perspectives, either by yielding more compact representations or enhanced interpretability.

In this context, \gls{tsp} has emerged as a framework for analyzing signals supported on a broad class of non-Euclidean, topological domains \cite{robinson2014topological, barbarossa2020topological}. Building upon tools from algebraic topology and differential geometry, \gls{tsp} lends itself to the modeling and dynamic analysis of multi-way relations and signals evolving across multiple scales. This framework has attracted increasing attention in recent years, driven the seminal works of Barbarossa and collaborators \cite{barbarossa2020topological, barbarossa2020topological2}. Their contributions have been key to developing processing approaches based on the combinatorial Hodge Laplacian, a generalization of the graph Laplacian for topological spaces rooted in Hodge theory, which has then been extended to other operators acting on topological domains \cite{bianconi2021topological, calmon2023dirac}.

Broadly speaking, \gls{tsp} generalizes \gls{gsp} by
accommodating for a wider range of domains and the associated signal representations. From the domain perspective, \gls{tsp} extends graphs to more general topological domains such as \emph{simplicial complexes} \cite{barbarossa2020topological}, \emph{cell complexes} \cite{sardellitti2021topological}, and other topological objects \cite{sardellitti2025topological, battiloro2024tangent}, which can explicitly and elegantly encode multi-way relationships in the data. From the signal perspective, \gls{tsp} enables the analysis of higher-order signals defined not only on nodes, but also on edges, triangles, and, more generally, topological elements of arbitrary dimension \cite{barbarossa2020topological, isufi2025topological, schaub2020random}.

From a theoretical standpoint, \gls{tsp} has rapidly expanded, extending a wide range of signal processing tools to general topological domains. Recent developments include, among others, the definition of the \gls{tft} and topological frequencies \cite{barbarossa2020topological}, the introduction of topological convolutional filters \cite{yang2022simplicial}, and early efforts to formulate probabilistic models for higher-order signals on complex domains \cite{sardellitti2023probabilistic, yang2023hodge, navarro2026stationarity}. Furthermore, the development of \gls{tsp} has contributed to the emergence of the rapidly growing field of \gls{tdl}, a branch of geometric machine learning related to the study of architectures that respect and leverage the topological backbone of the data \cite{ebli2020simplicial, yang2023convolutional, battiloro2024generalized, maggs2023simplicial, papamarkou2024position}.\\

\subsection{Applications of TSP and Limitations}
Building on the above theoretical foundations, processing based on the Hodge Laplacians and related operators offers a powerful framework for studying communication patterns and multi-way interactions in complex datasets defined over non-Euclidean domains. Initial applications of \gls{tsp} have begun to emerge in several areas, including for instance the modeling of water distribution networks \cite{cattai2025physics,cattai2025leak}, brain imaging \cite{bispo2025learning, Santoro2025EdgeLaplacians, sardellitti2025topological, bispo2026multimodal}, and the analysis of ocean trajectories \cite{schaub2020random}, among others \cite{isufi2025topological,bick2023higher,battiston2020networks}.

In parallel with these developments, a number of authoritative reviews have provided unifying perspectives on \gls{tsp} and related research areas, approaching the field from complementary viewpoints. Existing works range from signal-processing–driven reviews \cite{schaub2021signal, schaub2022signal, barbarossa2020topological2} to perspectives rooted in topological machine learning \cite{isufi2025topological, papamarkou2024position, hajij2022topological}. Other reviews emphasize the geometry and topology of higher-order networks \cite{bick2023higher,abiad2026hypergraphs}, leverage physics- and dynamical systems–based approaches \cite{battiston2020networks}, or present topological concepts and objects from an algebraic yet application-friendly perspective \cite{hoppe2025don}. Together, these works have shaped the conceptual landscape of \gls{tsp} and opened the avenue for future research endeavors.

A recurring message across these reviews is that translating \gls{tsp} theory and related frameworks into data-driven applications remains a central challenge \cite{bernardez2026topological}. While substantial progress has been made at the theoretical level, and despite the growing interest from a wide research community, the application of \gls{tsp} to real-world data remains largely exploratory and practical implementations continue to lag behind for several reasons. First, the notions of algebraic topology and differential geometry necessary to access the framework may represent a significant barrier for non-experts. Second, higher-order topological domains and signals are rarely both directly available in the data. Indeed, most non-Euclidean datasets are provided in the form of graphs, and observed signals are more typically defined on nodes rather than on edges or higher-order structures. Therefore, higher-order signals or domains are often missing in the initial data, even though the benefits of \gls{tsp} are best realized when both are jointly processed. These mismatches between modeling components and empirical data significantly hinder the use of \gls{tsp}-based methods in applied settings. Addressing these aspects has the potential to create a virtuous cycle, in which advances in empirical applications inform the development of new methodological tools, ultimately strengthening the impact of \gls{tsp} across disciplines.

\subsection{Positioning and Contributions of This Work}
Motivated by this open challenge, we provide a tutorial on \gls{tsp}, adopting an application-oriented perspective and focusing on signal processing based on the Hodge Laplacian operator for simplicial complexes. Our objective is to provide readers with the conceptual and practical tools required to apply \gls{tsp} effectively by operating on two complementary fronts. First, by clarifying the foundational algebraic topology and signal processing concepts underlying any \gls{tsp}-based analysis, to help readers approach the field with a solid understanding of its theoretical foundations. Second, by addressing the challenges of leveraging \gls{tsp} under suboptimal conditions, for instance when only lower-order domains or signals are initially available in a dataset.

To this purpose, we introduce the core \gls{tsp} and related algebraic topology concepts through a signal-processing lens, highlighting how these abstract notions can be translated into practical tools for analyzing real-world data. Guided by George Box’s adage that ``all models are wrong, but some are useful,'' we emphasize how defining or deriving domains and signals on higher-order structures can reveal dynamical and topological features that are otherwise hidden in node-centric analyses. We also present a real-world use case demonstrating the applicability of these concepts, using brain imaging data. We choose this specific application since brain imaging data have a structure which makes them suitable for \gls{gsp}-based analysis, and has made them the target of some \gls{tsp} applications \cite{Santoro2025EdgeLaplacians,sardellitti2025topological,bispo2025learning, bispo2026multimodal}. 

\subsection{Paper Structure}
In Sec.~\ref{sec:Background_TSP}, we introduce the fundamental \gls{tsp} preliminaries and concepts, highlighting not only their theoretical underpinnings but also their interpretation and use in practical applications. Then, in Sec.~\ref{sec:Domain_and_Edge_Signals}, we present the challenges involved in constructing higher-order domains and signals from simpler data in a manner that is both compatible with \gls{tsp} and instrumental for extracting useful information for downstream analysis. In particular, this section shows how an informative edge signal can be obtained from nodal time series through simple transformations, providing a concrete example of how processing the original data appropriately can produce interpretable higher-order signals for analysis with \gls{tsp} tools. In Sec.~\ref{sec:Application}, we demonstrate the application potential of \gls{tsp} by leveraging the previously defined edge signal with a case study scenario on brain imaging data. In Sec.~\ref{sec:Discussion}, we summarize the main topics and potential avenues for future research, to conclude with Sec.~\ref{sec:conclusions}.

In various sections of the paper, light blue boxes are used to provide examples or application-oriented insights on more theoretical concepts. Furthermore, a few technical remarks are included in selected footnotes and may be omitted by the reader without affecting understanding of the main text.

%% file: Main_text/Sections/1_Background_TSP.tex
\section{An Application-Oriented Tutorial on \gls{tsp}}
\label{sec:Background_TSP}

\noindent We start with a tutorial on the fundamental principles underpinning \gls{tsp} with the Hodge Laplacian. The definitions and ideas presented here draw from elementary notions of algebraic topology and differential geometry, and require only a foundational background in linear algebra and signal processing. We demonstrate how theoretical notions translate into practical applications bringing in intuition and illustrations.

A useful way to approach \gls{tsp} is to view it as a paradigm built on two complementary components: \emph{(i)} a topological domain or space; \emph{(ii)} a signal living on top of it. Therefore, in this section we follow the same dual structure. Special emphasis is placed on the discrete algebraic topology-based domains as they pose a common initial barrier for non-expert readers, yet they play a crucial role in the subsequent processing. We then introduce the concept of higher-order signals living on topological domains. Consistent with the application-oriented nature of this paper, after presenting topological signals in a general form, we concentrate mainly on signals living on the \emph{edges} of a network, given their central importance for the majority of applications. We proceed by showing how the properties of a domain can inform signal processing. In particular, we discuss notions such as smoothness, spectral decomposition, and the Hodge Laplacian-based Fourier transform. Finally, we draw parallels between \gls{tsp} and \gls{gsp}, highlighting how new \gls{tsp} ideas extend and relate to established \gls{gsp} concepts.

\subsection{Notation}
Across the paper, the following conventions are used. We use bold capital letters to denote matrices and bold lowercase letters for column vectors. The entries of a matrix $\ma{X}$ and a vector $\vc{x}$ are denoted by $\left[\ma X \right]_{ij}$ and $x_i$, respectively. The notation $(\cdot)^\top$ denotes transpose. We write $\operatorname{diag}(\lambda_1, \dots, \lambda_E)$ for a diagonal matrix with entries $\lambda_1, \dots, \lambda_E$, and $\mathrm{blkdiag}(\ma \Lambda_1, \dots, \ma \Lambda_k)$ for a block-diagonal matrix with blocks $\ma \Lambda_1, \dots, \ma \Lambda_k$. $\ma{I}$ denotes the identity matrix. We denote by $\|\vc{x}\|_2$ the $\ell_2$-norm of a vector $\vc{x}$, defined as $\|\vc{x}\|_2 = \sqrt{\sum_i |x_i|^2}$, and by $\|\ma{X}\|_F$ the Frobenius norm of a matrix $\ma{X}$, defined as $\|\ma{X}\|_F = \sqrt{\sum_{i,j} |X_{ij}|^2}$. We use calligraphic capital letters for sets and plain capital letters for their cardinalities; e.g., $\mathcal{V}$ denotes the set of vertices and $V$ its cardinality. Unless otherwise specified, $E$ and $T$ denote the number of edges and triangles, respectively. The operators $\operatorname{im}(\cdot)$ and $\operatorname{ker}(\cdot)$ denote the image and kernel of a matrix, respectively, and $\oplus$ denotes the direct sum of vector spaces. Furthermore, $\mathbb{E}$ denotes expectation. For signals of any order, we use $\vc{x}^{(i)}$ to indicate a signal of order $i$.

\subsection{The Domain Ingredients: Simplices and Simplicial Complexes}
A natural starting point is the definition of a simplex that is a topological object well-suited to modeling high-order relationships.
\subsubsection{Simplex} 
Given a finite set of points, or vertices, denoted by $\mathcal{V} = \{v_1, v_2, \dots, v_{N}\},$ where $N$ is the total number of vertices, an \emph{abstract $k$-simplex} $\sigma^k$ is an unordered set of the $k+1$ vertices:
\begin{equation}
    \label{eq:simplex}
    \sigma_i^k = \{v_{i_1}, v_{i_N}, \dots, v_{i_k}\} \subseteq \mathcal{V},
\end{equation}
\noindent with indices satisfying $1 \leq i_j \leq N$ and $v_{i_j} \neq v_{i_m}$ for every $j \neq m$, as the vertices must be distinct. The \emph{dimension} or \emph{order} of a simplex is defined as one less than the number of its vertices. 

\begin{figure*}[!t] 
  \centering
  \includegraphics[width=.85\linewidth]{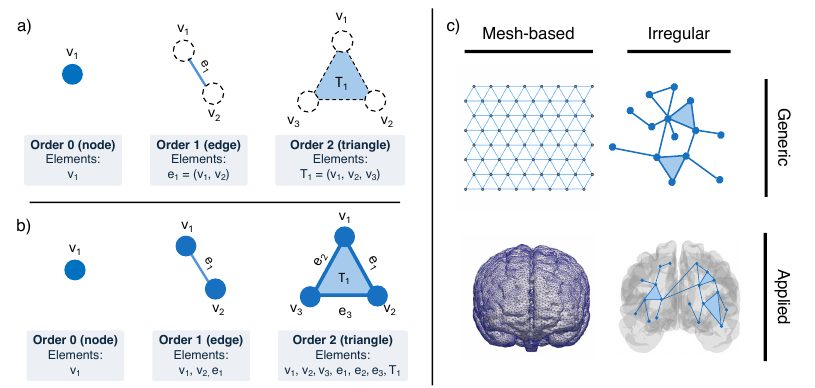} 
  \caption{\textbf{Simplices and Simplicial Complexes.} 
    a) Simplices of different orders depicted  as geometric objects in the plane. b) Corresponding simplicial complexes obtained by including all faces of each simplex. 
    c) Different types of simplicial complexes (top) and an example application for brain imaging modeling (bottom). 
    For the brain mesh, the neighborhood relationships are determined by the cortical surface. 
    In contrast, the bottom-right simplicial complex can be built from a graph of white matter connections between brain regions derived from diffusion imaging.}
  \label{fig1:Simplices_and_Simplicial_Complexes}
\end{figure*}

Abstract simplices are purely combinatorial entities and need not be associated with any geometric space. Nevertheless, it is often helpful to visualize them via their geometric interpretation, where $0$-simplices correspond to points, $1$-simplices to edges, $2$-simplices to triangles, and so on, as shown in Fig.~\ref{fig1:Simplices_and_Simplicial_Complexes}. A simplex can indeed be viewed as a geometric object when considering its so-called \emph{realization} in a metric space, such as $\mathbb{R}^M$ for some real value $M$. This corresponds to assigning each vertex of the simplex to a value in that space, plus a few additional properties to be satisfied\footnote{In particular, associating an embedding to a simplex $\sigma^k$ into $\mathbb{R}^M$ requires the mapping function $\phi: \sigma^k \rightarrow \mathbb{R}^M$, which assigns each vertex of $\sigma^k$ to a value in $\mathbb{R}^M$, to be an \emph{affine embedding}, a condition ensuring that the simplex does not degenerate by collapsing along any of its dimensions. For more details see \cite{hatcher2001algebraictopology, nanda2016algebraic}.}. The important conceptual difference is that an abstract simplex describes only \emph{how} nodes are grouped, whereas an embedded simplex specifies the manner in which these connections occur \emph{within the space}. To foster intuition, throughout this work we refer to simplices and their geometric realizations interchangeably. However, the topological domains considered here are not restricted to geometric realizations. We therefore adopt a mild abuse of notation to simplify the exposition of concepts that would otherwise be difficult to visualize, while ensuring that all notions and results remain valid independently of any embedding.

\textbf{Remark: domain \emph{vs} signal} To leverage \gls{tsp} properly, the $k$-adic relationships modeled as a $k$-simplex is supposed to be present in the \emph{structure} of the dataset. Indeed, we assume that simplices capture stable, characteristic properties of the data, which influence the dynamics of a signal defined on top of them and its propagation across the domain.  Similarly to \gls{gsp}, the domain structure may, in principle, also vary over an additional dimension such as time (for a recent review covering the topic in \gls{gsp}, see \cite{yan2024signal}). Nevertheless, to avoid ambiguity and keep the focus on signal dynamics, in this work the topology of the domain is assumed to be fixed.

\begin{insightbox}{\textbf{Simplices for Modeling Relations in Data}}
    From an applied perspective, a node may represent an individual element of a dataset or a location where data can be measured, and a $k$-simplex can model a $k$-adic relationship among nodes. This relationship can correspond to a physical multi-way connection or to patterns inferred according to a suitable criterion (see Sec.~\ref{subsec:domain} for further discussion on the topic).
    For instance, in a dataset containing information from a social network, each user can be represented as a node, whereas each simplex can capture multi-user connections, such as groups of users that frequently exchange messages simultaneously or belong to the same group chat. 

    Moving to geometric simplices in applied settings, embeddings frequently arise naturally in deep learning contexts. One simple example is given by datasets represented as graphs endowed with feature vectors assigned to their nodes (i.e., $0$-simplices). In this case, associating to each vertex a vector of $M$ features is equivalent to defining a nodal embedding map $\phi: V(G) \rightarrow \mathbb{R}^M$, where $V(G)$ denotes the set of vertices of the graph $G$. This embedding allows each node to be associated with a point in $\mathbb{R}^M$, and consequently induces a geometric representation of the $1$-simplices as edges in that space. An example of how these concepts are used to build topologically informed neural networks can be found in \cite{maggs2023simplicial}.
\end{insightbox}

One pervasive idea in algebraic topology is that the properties of a $k$-simplex are determined by its relationships with simplices of adjacent orders, and in particular those that form it and those that contain it. These relationships are formalized through the notion of incidence and expressed in terms of faces and cofaces.

\subsubsection{Incidence, Faces, and Cofaces} 
Given an abstract $k$-simplex $\sigma$, we say that a $(k-1)$-simplex $\tau $ is \emph{incident} to $\sigma$ if $\tau \subset \sigma$, that is, if the set of vertices of $\tau$ is obtained by removing exactly one element from the vertex set of $\sigma$. In this case, $\tau$ is called a \emph{face}\footnote{Note that, depending on the context, faces of a $k$-simplex can refer both to $(k-1)$-incident simplices or to incident simplices of any order $0 \leq j \leq k-1$. In the latter case, incident simplices of order $(k-1)$ are called \emph{facets}.} of $\sigma$. A $k$-simplex $\sigma$ has exactly $k+1$ faces.
Similarly, a \emph{coface} of a $k$-simplex $\sigma$ is a $(k+1)$-simplex having $\sigma$ as a face.

For instance, the faces of an edge ($1$-simplex) connecting two nodes ($0$-simplices) are the nodes themselves. In addition, if an edge is part of a triangle ($2$-simplex), that triangle is a coface of the edge. A visual representation of these concepts is provided in Fig.~\ref{fig1:Simplices_and_Simplicial_Complexes}.

Equipped with simplices as building blocks and a notion of incidence, the next step is to define the domain that underlies the \gls{tsp} analysis of this work, a topological object composed of multiple simplices under the constraint of inclusion.
\subsubsection{Simplicial Complex} 
An \emph{abstract simplicial complex} $\mathcal{X}$ is a finite collection of simplices closed under inclusion of faces: given any simplex $\sigma \in \mathcal{X}$, then every face of that simplex must also belong to $\mathcal{X}$.

For instance, to obtain a simplicial complex of order $2$ from a simplex of the same order---an empty triangle or, equivalently, a \emph{hollow} $2$-simplex---it is necessary to consider the triangle itself and include all its edges and nodes. Examples of combinatorial and geometrical simplicial complexes are shown in Fig.~\ref{fig1:Simplices_and_Simplicial_Complexes}. 

Despite the abstract definition, in real applications simplicial complexes can exhibit a broad range of forms. For example, any graph is by definition a simplicial complex of order $1$, as it includes nodes and edges between them. Furthermore, starting from a graph, a simplicial complex of order $2$ can be obtained by filling the triangles between all triplets of existing edges, which are also referred to as $3$-cliques. This type of simplicial complex is called \emph{clique complex} (or \emph{flag complex}) \cite{CHONG2021105466}. Note that a collection of geometric simplices defines a \emph{geometric simplicial complex} provided that the intersection of any two simplices is a common face of both, considering the empty set a face of any simplex \cite[Sec.~2]{munkres2025elements}. For instance, any two triangles may share only a common vertex or a common edge, but they cannot intersect in a more arbitrary way, such as crossing through the interior of one another. A mesh resulting from the triangulation of a surface is an example of a geometric simplicial complex.

A simplicial complex $\mathcal{X}$ is said to be \emph{weighted} if a weight $\alpha_i$, typically in $\mathbb{R}$, is assigned to each simplex $\sigma_i \in \mathcal{X}$. Intuitively, a weight quantifies the strength of the relationship encoded by the simplex, such as in graphs the weight of an edge represents the strength of the binary connection it models. For simplicity, this work considers only unweighted simplicial complexes.

\begin{insightbox}{\gls{tsp} of Signals on Meshes}
    The flexibility of \gls{tsp} stems from its ability to handle signals defined on arbitrary networks and non-Euclidean supports. In settings where the underlying domain exhibits additional geometric regularity, as in meshes, classical tools from vector calculus and differential geometry, together with discretization frameworks such as the finite element method, can be employed to exploit the geometric structure of the domain. Moreover, different topology-aware operators may be defined on the same domain, and their properties may vary depending on whether they arise from purely combinatorial constructions or from geometry-aware discretizations (see, e.g. \cite{ribando2024combinatorial}). In any case, \gls{tsp} remains a versatile framework for analyzing signals on meshed domains, particularly when the focus lies on the relational structure of the data. In addition, regular meshes provide an intuitive setting for illustrating fundamental \gls{tsp} concepts, as their geometry naturally supports domain decompositions amenable to visual interpretation, as shown in Fig.~\ref{fig:Laplacian_evecs}. 
\end{insightbox}

One remarkable feature of simplicial complexes arises from their deep connection with algebraic topology. By modeling domains with simplicial complexes, one can capture global topological properties of the network, which are strongly connected to the presence of \emph{holes} and \emph{cavities}. Importantly, these holes are not local features of individual simplices, but rather emergent properties of their overall arrangement forming the domain. Informally, a hole arises when a collection of simplices forms a closed structure that is not covered by simplices of higher dimension. In particular, $1$D holes correspond to cyclic structures formed by edges that are not filled by $2$-simplices. As a trivial example, a loop of edges forming the boundary of a hollow triangle constitutes a $1$-dimensional hole: although the edges form a closed cycle, the absence of the corresponding $2$-simplex prevents the cycle from being the boundary of a filled region. Beyond this case, holes can also emerge from more complex configurations of simplices forming nontrivial cycles. An instance of a mesh with a hole is illustrated in Fig.~\ref{fig:Laplacian_evecs}.

\textbf{In the language of topology: holes as homology classes} Holes capture essential topological features of a space. Characterizing these features systematically is the goal of homology theory, which provides algebraic invariants that distinguish topologically different structures \cite{nanda2016algebraic, hatcher2001algebraictopology}. In this context, holes correspond to homology classes, i.e., equivalence classes of cycles encoding global topological properties of the domain \cite{hatcher2001algebraictopology}.
\footnote{In particular, holes correspond to topological features that are invariant under homeomorphisms of the space.} When the domain is modeled as a simplicial complex, these features can be computed via the Hodge Laplacian, as discussed in Sec.~\ref{subsec:Hodge}.

\begin{insightbox}{To Fill or Not to Fill Holes in Real Datasets}
{
    In applications, one of the main modeling challenges when deriving a simplicial complex from lower-order data, for instance when estimating a simplicial complex of order $2$ starting from a graph, is determining how and where to “fill the holes” in the domain.
    
    In some cases holes may appear naturally due to the structure of the underlying support, for instance when the complex is built using the geometric approaches discussed in Sec.~\ref{subsec:domain}. Alternatively, they can be deliberately introduced to highlight specific topological features that are known \emph{a priori} or to restrict signal propagation \cite{ghosh2018topological}.  In the case of measuring ocean's current trajectories \cite[Sec.~5.2]{schaub2020random}, the authors build the a simplicial complex by discretizing a geographical area underlying the sea using a hexagonal grid, and introducing a central hole in the mesh to represent the presence of the Madagascar island, because it acts as an obstacle to the trajectories.
    
    While it is relatively straightforward to leverage such priors for meshes or discretized domains, the problem becomes more subtle on irregular networks. In this context, it is crucial to understand how different construction strategies influence the representation of the domain and the downstream analyses. Importantly, from a signal processing perspective, the placement of holes strongly impacts the amount and nature of information that can be extracted from the signal of interest. A more detailed discussion of the implications of this choice is provided in Sec.~\ref{subsec:domain}.
}
\end{insightbox}

\subsubsection{Why Simplicial Complexes?}
A question naturally arising in the context of applications is why one might choose to model a domain using simplicial complexes, given the stringent inclusion criteria that make this framework relatively restrictive. At first glance, a more intuitive alternative might be to employ more familiar structures, such as hypergraphs, which offer greater flexibility and do not require the additional machinery of algebraic topology \cite{bick2023higher, battiston2020networks, abiad2026hypergraphs} (for an introduction to hypergraph signal processing, see \cite{zhang2019introducing, schaub2021signal}). 

While the above argument may apply in certain contexts, simplicial complexes often offer advantages that justify their use for domain modeling. First, thanks to the inclusion property, a simplicial complex naturally provides a \emph{hierarchical representation} of higher-order interactions and multi-way relations on the data, which is preserved during processing. Second, and perhaps most importantly, the algebraic representations of simplicial complexes provide powerful insights into the structure of the data. Indeed, simplicial complexes and related topological objects can describe both \emph{local} and \emph{global} properties of the domain they model. Global properties are reflected in the presence of (higher-dimensional) holes in the domain, while local properties arise from the neighborhoods relationships between simplices. Most conveniently, all these features are elegantly encoded in the combinatorial Hodge Laplacian operator, which will be introduced in the Sec.~\ref{subsec:Hodge}. As a consequence, from a signal processing standpoint, this framework enables the study and decomposition of signals of any order according to global or local features, providing a more nuanced understanding of the signal or, at the occurrence, a more compact representation. Thus, the strict inclusion criteria of simplicial complexes are ultimately the price for accessing powerful algebraic tools, gaining elegant insights, and simplifying computations. 

These topics are explored in more detail in the following paragraphs, through the boundary and Hodge Laplacian operators.

\subsection{An Arrow Is All It Takes: Oriented Simplices, Boundary Operators, and Topological Signals}
\label{subsec:hodge_boundary}
When processing higher-dimensional signals on simplicial complexes, it is necessary to assign an orientation to each simplex to ensure coherent computations. This concept is closely linked to the ordering of its vertices.

\subsubsection{Oriented Simplex}
A $k$-simplex becomes \emph{oriented} upon specifying an ordering of its vertices.
Note that this ordering is arbitrary, but must be fixed consistently for bookkeeping and computational purposes. 

For instance, a simplex of order $0$ ({e.g.}, the vertex $v_1$) has only one trivial orientation, because a single element cannot be reordered. In contrast, the 1-simplex $\{v_1,v_2\}$  becomes oriented once the link between $v_1$ and $v_2$ is assigned a direction, then the link can be represented by an arrow. This concept is illustrated in Fig.~\ref{fig:ExampleIncidenceMatrices}.
Accordingly, there are two possible orientations: the arrow may have tail in $v_1$ and tip in $v_2$, denoted as $e_1=[v_1, v_2]$, or vice versa, denoted as $[v_2, v_1]$. Similarly, the $2$-simplex $t_1=\{v_1, v_2, v_3\}$ becomes oriented when provided with one of the two possible ordering $[v_1, v_2, v_3]$ or $[v_1, v_3, v_2]$, corresponding to an arrow rotating clockwise or counter-clockwise along the vertices. Note that the first configuration is equivalent to $[v_2, v_3, v_1]$ and $[v_3, v_1, v_2]$ because the orientation does not depend on when the rotating arrow starts. More formally, an orientation corresponds to an equivalence class of vertex orderings, and two orientations are equivalent if they differ by an even permutation. For a physical intuition, one can think of the two orientations of $2$-simplices as an arrow pointing upwards or downwards relative to the triangle plane surface. In general, oriented simplices of any orders have two opposite orientations.

One common and convenient choice, also adopted in this work, is to assign the orientation defined by the numerical or lexicographic order of vertices. Accordingly, an edge between $v_1$ and $v_2$ becomes $[v_1, v_2]$, with tail in $v_1$ and tip in $v_2$. Similarly, the $2$-simplex $[v_1, v_2, v_3]$ is oriented with a rotating arrow following the vertex numerical ordering, as in Fig.~\ref{fig:ExampleIncidenceMatrices}.

\textbf{Remark: orientation \emph{vs} direction} In the simplicial complexes considered in this work, arrows indicate an \emph{orientation}, not a direction. That is, a signal defined on an oriented edge can propagate in both directions; the orientation only establishes a consistent sign convention for each edge. The same principle applies to higher-order signals.

\subsubsection{Boundary Operators}
Given an unweighted oriented simplicial complex $\mathcal{X}$, the $k$-th \emph{boundary operator} encodes the incidence relation between simplices of order $k$ and order $k-1$, and is algebraically represented by the incidence matrix $\mathbf{B}_k$. Its entries are defined as follows:
\begin{equation}
    \label{eq:boundary_op}
    [\mathbf{B}_k]_{i, j} =
    \begin{cases}
        0, & \text{} \sigma^{k-1}_i \text{not a face of } \sigma^{k}_j, \\
        1, & \text{} \sigma^{k-1}_i \text{face of } \sigma^{k}_j \text{ and orientations agree}, \\
        -1, & \text{} \sigma^{k-1}_i \text{face of } \sigma^{k}_j \text{ and orientations disagree}.
    \end{cases}
\end{equation} 
Each row of $\mathbf{B}_k$ represents a specific $k$-simplex in $\mathcal{X}$, whereas the columns are the simplices of order $k+1$. An illustrative example is presented in Fig.~\ref{fig:ExampleIncidenceMatrices}. For a simplicial complex of order $2$, with $V$ vertices, $E$ edges, and $T$ triangles, the boundary operator mapping $2$- to $1$-simplices is encoded by the boundary matrix $\mathbf{B}_1 \in \mathbb{R}^{V \times E}$, with each row corresponding to one node and each column to an edge of the complex. For instance, the column of $\mathbf{B}_1$ related to edge $e_1$ has per entry $0$ if the node of the corresponding row is not a face of $e_1$, $-1$ if the node is the tail of $e_1$, and $+1$ if the node is the tip of $e_1$. Similarly, the boundary matrix $\ma B_2 \in \mathbb{R}^{E \times T}$ captures the incidence relationships between edges and triangles: the column of $\ma B_2$ associated to the triangle $t_1$ has per entry $0$ if  $e_1$ is not a face of $T_1$, $-1$ if  $e_1$ is part of $t_1$ but its orientation is not coherent with the one of $t_1$ (i.e., the arrow of the edge and the rotating arrow of the triangles point in opposite directions), and $1$ if  $e_1$ is part of $t_1$ and the orientations are coherent. These concepts are illustrated in Fig.~\ref{fig:ExampleIncidenceMatrices}. 

A fundamental property of these operators, deeply rooted in topology, is that the boundary of a boundary is zero. Algebraically, this is expressed as
\begin{equation}
    \label{eq:boundary_of_boundary_is_zero}
    \bk \bkpu = \mathbf{0}, \quad \forall k \in 1, \dots, K.
\end{equation}
This identity follows directly from the definition of the boundary operators and the consistency of simplex orientations (see also \cite{barbarossa2020topological, hatcher2001algebraictopology, nanda2016algebraic}).

\subsubsection{Higher-Order Signals on Oriented Simplicial Complexes}
Given an oriented simplicial complex $\mathcal{X}$ of order $K$, where the number of simplices in each dimension $k$ is $N_k$, a $k$-th order simplicial signal $\vc{x}^{(k)}$ is a mapping from the $k$-simplices of $\mathcal{X}$ to real values. Algebraically, it admits a coordinate vector representation in $\mathbb{R}^{N_k}$, where each entry $x_i^{(k)}$ corresponds to the value associated with the simplex $\sigma^k_i \in \mathcal{X}$. 

In practice, a $0$-order signal lives on the $N$ nodes of a simplicial complex and can be represented by a vector in $\mathbb{R}^{N}$; a $1$-order signal on the edges corresponds to a vector in $\mathbb{R}^{E}$; more generally, a $k$-order signal, which lives on $k$-dimensional simplices, is represented by a vector in $\mathbb{R}^{N_k}$.
Note that a simplicial complex of order $K$ allows the analysis of signals defined on simplices of order $K$ or lower. For instance, given a simplicial complex of order $2$, it is possible to analyze signals living either on its nodes, edges, or triangles. This setting is the most common in practical applications. 

Moreover, in principle the \gls{tsp} framework also permits to jointly analyze signals on simplices of different orders through the use of the Dirac operator. We refer to \cite{bianconi2021topological, calmon2023dirac} and related works from Bianconi and colleagues for a detailed treatment of this topic.

\begin{insightbox}{\textbf{Higher-Order Signals in Real Datasets}}{
    Most commonly, higher-order signals arise naturally in a variety of application domains in the form of edge signals. An example is provided by communication data in social networks, where an edge signal may represent the number of messages exchanged between two users, with each message being uniquely associated with a sender and a receiver. Other edge-level data include traffic flows and mobility data \cite{schaub2020random,ghosh2018topological}, patient flows between hospitals \cite{gebhart2021go}, cell differentiation trajectories \cite{cheng2025phlower}, ocean velocity \cite{yang2023hodge}, or financial transactions \cite{yang2023hodge}.
    
    Despite these examples, most datasets comprise signals that are inherently nodal, as they derive from measurements collected at pointwise locations. In such settings, edge or other higher-order signals are not directly available; however, it can be meaningful or useful to derive such signals from nodal observations, particularly when the nodal measurements are assumed to reflect or be related to processes occurring at a higher-order level. This approach has been applied, for instance, in a few works on brain imaging data \cite{bispo2026multimodal, Santoro2025EdgeLaplacians}; we further address it and present a concrete application in Sections \ref{sec:Domain_and_Edge_Signals} and \ref{sec:Application}, respectively.
    }
\end{insightbox}

\begin{figure*}[!t]
  \centering
  \includegraphics[width=.92\linewidth]{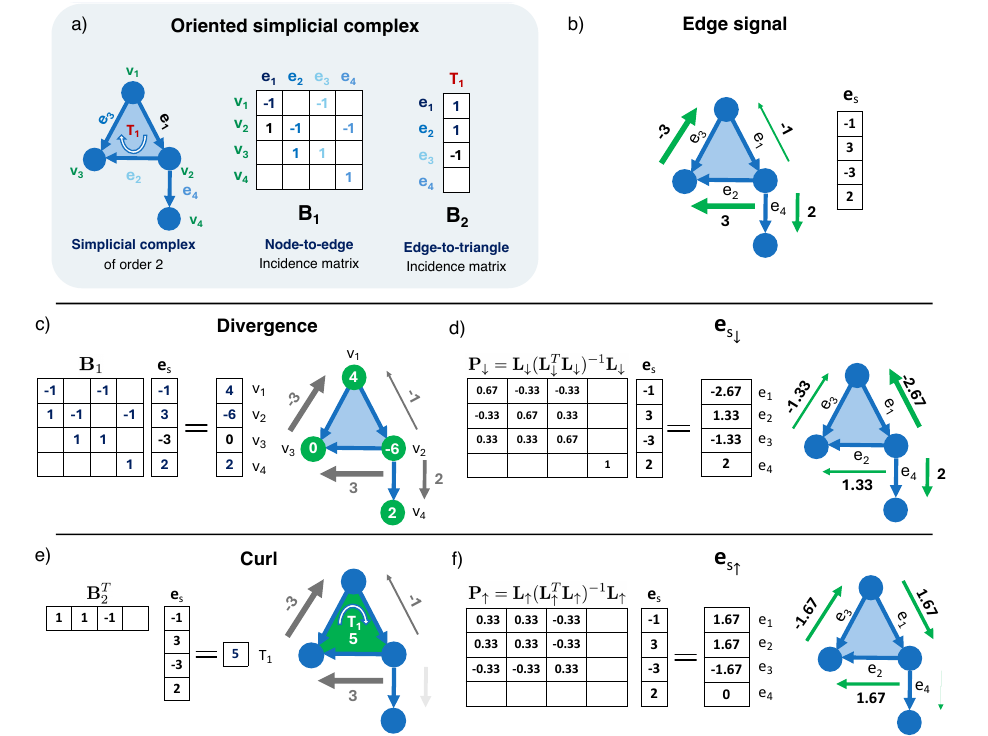} 
  \caption{\textbf{Oriented Simplices, Boundary Matrices, and Operations on Edge Signals.}
  a) Example of oriented simplicial complex of order $2$ and its associated boundary matrices. Empty entries indicate zero. b) Edge signal living on the complex. The sign of each entry indicates whether the signal is aligned or anti-aligned with the reference orientation of the corresponding edge, whereas the thickness of the arrow is proportional to the signal magnitude. c) Divergence, obtained by computing the difference between inflow and outflow at each node. e) Curl, obtained by computing the divergence-free flows circulating around the $2$-simplex. d), f) Irrotational and solenoidal components of the edge signal, obtained as the linear projection of $\vc{e}_s$ into the space of the lower and the upper Laplacian, respectively. Matrix entries are shown truncated to two decimal places for visualization purposes, whereas computations are performed in full precision. A useful remark is that $\vc e_{s\downarrow}$ and the divergence signal contain the same type information, as well as the $\vc e_{s\uparrow}$ and the curl signal, although represented in spaces of different order. Note that the signal in this image does not present any harmonic component, as the only $2$-simplex is included in the simplicial complex (i.e., the only triangle is ``filled") and there are no residual circulations.}
  \label{fig:ExampleIncidenceMatrices}
\end{figure*}

While the previous definition of higher-order signals is quite general, we restrict our attention to signals that are most effectively analyzed within \gls{tsp}, namely those compatible with orientation. This aspect marks a key distinction from \gls{gsp} and entails a more refined interpretation of signal sign. As an example, consider an oriented edge $e_1=[v_1, v_2]$. In this setting, a positive signal on $e_1$ can be interpreted as a physical or abstract quantity flowing from $v_1$ to $v_2$, while a negative signal may be indicate the same quantity flowing in the opposite direction. 
If we change the convention and set  $\tilde{e}_1=[v_2, v_1]$, the signal sign should be inverted to describe the same flow respect to $\tilde{e}_1$. Analogous reasoning can be extended to order 2: oriented signals on $2$-simplices can be interpreted as circulations, representing quantities that rotate around each triangle either clockwise or counter-clockwise depending on their sign with respect to the chosen reference orientation. These concepts are illustrated in Fig.~\ref{fig:ExampleIncidenceMatrices}, and remain valid for signals of any order. 

From now on, we assume that all simplices and simplicial complexes are oriented. We also omit the qualifier “oriented” for higher-order signals whenever it can be clearly inferred from the context. Furthermore, we use the terms “topological signals” and “higher-order signals” almost interchangeably, with the distinction that topological signals include all orders starting from 0, whereas higher-order signals refer to signals starting from order 1.

\begin{insightbox}{Caveats When Grouping Higher-Order Signals}{
    It is often useful to aggregate signals according to specific grouping criteria; e.g. when analyzing road networks to study traffic trends, streets may be grouped by geographic area or city, which can answer whether collections of roads connecting particular regions exhibit similar traffic patterns. For such aggregations, the sign of the signal encodes its alignment or misalignment with respect to a chosen reference orientation. 
    
    Let us consider a rail road connecting the cities of Milan and Lausanne, conceptualized as an edge \emph{Milan-Lausanne} oriented from Milan (arrow tail) to Lausanne (arrow tip). A positive signal on this edge corresponds to flow of trains aligned with this orientation, whereas a negative signal indicates an opposite flow, directed to Milan. When aggregating edges, for instance by defining a country-level edge \emph{Switzerland-Italy} from regions in Switzerland (tail) to regions in Italy (tip), the sign of the original inter-city edge signal must be inverted prior to aggregation, to ensure consistency with the new reference orientation.

    This principle extends to signals of any order: aggregation must account for orientation parity between the simplex and the induced group structure, with a sign correction when the two are mismatched.
}
\end{insightbox}

\textbf{In the Language of Topology: Chains and Cochains} 
The term \emph{cochains} is often used in more topology-informed literature to denote higher-order signals living on oriented simplices; for example, flows on the edges of a graph can be identified with a $1$-cochain. This notion directly stems from standard representations in algebraic topology. Given a simplicial complex $\mathcal{X}$ of finite dimension $K$, one can construct a vector space by considering all formal linear combinations of $k$-simplices with coefficients in $\mathbb{R}$, for $k \leq K$. This vector space is denoted by $C_k(\mathcal{X}, \mathbb{R})$, and its elements are called \emph{$k$-chains}. In this context, a $k$-order signal can be viewed as a cochain, i.e., a linear function from $C_k(\mathcal{X}, \mathbb{R})$ to $\mathbb{R}$. Therefore, a cochain is an element of the dual space of $C_k(\mathcal{X}, \mathbb{R})$, known as $C^k(\mathcal{X}, \mathbb{R})$. In more concrete terms, while chains describe the combinatorial structure of the complex acting as geometric ``building blocks", cochains can be seen as signal values on them (see also \cite{hatcher2001algebraictopology}). 

\textbf{In the Language of Differential Geometry: Differential $k$-Forms.}
Beyond their interpretation as cochains, signals of order $k$ living on oriented simplices can be viewed as discrete $k$-forms on simplicial complexes, i.e., the discrete analogue of $k$-differential forms on manifolds. Importantly, this deep connection underpins most theoretical aspects of topological signals and \gls{tsp} based on the combinatorial Hodge Laplacian. An introductory yet rigorous treatment of the field can be found in \cite{crane2018discrete}.

\subsubsection{Analyzing Edge Signals With Boundary Matrices: Divergence and Curl in Action}
Boundary matrices can be treated as linear operators acting on topological signals. Here, we provide an illustrative example of their use by considering a network of pipes with a fluid flowing through them; i.e., a simplicial complex of order $2$, with the fluid flow as edge signal.

A first quantity of interest is the amount of fluid entering or leaving each node, which depends on the balance of flows over the edges incident to that node. This information is captured by the edge signal together with the incidence matrix $\mathbf{B}_1$. Specifically, $\mathbf{B}_1$ encodes both edge–node incidence and edge orientations, enabling the computation of net inflow or outflow at each node from the signed edge values. The resulting nodal quantity corresponds to the discrete \emph{divergence} of the edge signal and can be computed for each node $i$ of the simplicial complex as follows:
\begin{equation}
    \label{eq:divergence}
    [\operatorname{div}(\vc x^{(1)})]_i = [\mathbf{B}_1 \vc x^{(1)}]_i= \sum_{e \in \text{in}(i)} x^{(1)}_e - \sum_{e \in \text{out}(i)} x^{(1)}_e ,
\end{equation}
where the two sums involve the edge flows entering and exiting node $i$, respectively. In practice, non-zero divergence indicates a net imbalance between outgoing and incoming flow at a given node, corresponding to a  sink (divergence $>0$) or a source (divergence $<0$).

Another relevant aspect of the example network concerns the presence of circulations, which correspond to flow rotating around triangles of the simplicial complex. Note that each node in a triangle experiences equal incoming and outgoing flow; therefore, by construction these rotations do not contribute to the divergence. This quantity is known as discrete \emph{curl}, and for each triangle $t$ is defined as follows:
\begin{equation}
    \label{eq:curl}
    [\operatorname{curl}(\vc x^{(1)})]_t = [\mathbf{B}_2\transpose \vc x^{(1)}]_t = \sum_{e \in \partial t} \sigma_{t,e} \, x^{(1)}_e ,
\end{equation}
where ${e \in \partial t}$ indicates that the sum is computed over the edges on the border of triangle $t$, and $\sigma_{t,e} = \pm 1$ indicates the sign, depending on whether the orientation of edge $e$ agrees ($+1$) or disagrees ($-1$) with the one of the triangle. An illustration of these calculations is shown in Fig.~\ref{fig:ExampleIncidenceMatrices}.

Notably, the above divergence and curl operators provide discrete versions of the classical divergence and curl from vector calculus, adapted to signals defined on simplicial complexes \cite{crane2018discrete}.

\begin{figure*}[t]
    \centering
    \includegraphics[width=.92\textwidth]{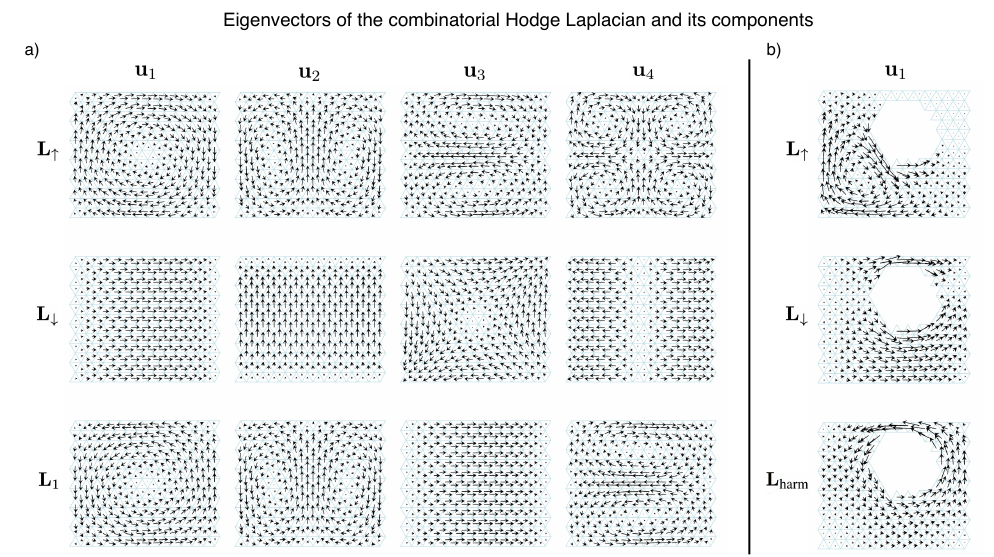}
    \caption{\textbf{Eigenvectors of the Laplacian of a regular mesh}. a) First eigenvectors of $\mathbf{L}_{1\uparrow}$, $\mathbf{L}_{1\downarrow}$, and $\mathbf{L}_{1}$ are shown. The underlying mesh is composed of edges with the same unit weight, and all triangles are included in the simplicial complex.
    Note that the eigenvectors are not visualized directly at the level of the edges, but rather as vectors at the barycenter of the mesh triangles to facilitate visual interpretation. The conversion is performed using a linear interpolation approach using Whitney basis functions (see \cite{barbarossa2020topological2} for details). Eigenvectors are displayed from left to right in order of increasing eigenvalues, which correspond to increasing frequency content of the signals. In particular, for $\mathbf{L}_{\uparrow}$ (first row), eigenvectors with larger eigenvalues exhibit increasingly complex rotational patterns, with a higher number of vortices as the frequency increases.
    On the other hand, the first eigenvectors of the $\mathbf{L}_{\downarrow}$ (second row) are associated with smooth oscillating gradients of edge signals.
    b) First eigenvector of different components of the Hodge Laplacian associated to a mesh with one hole. $\mathbf{L}_{\text{harm}}$ indicates the only eigenvector of the kernel of $\mathbf{L}_1$. Note that the mesh has only one harmonic eigenvector, since each of them is associated to a distinct hole in the space. The harmonic eigenvector captures rotational patterns around the hole, while the solenoidal and irrotational eigenvectors encode rotational patterns outside the hole and gradient-like variations along edges, respectively.}
    \label{fig:Laplacian_evecs}
\end{figure*}

Moreover, if the network contains hollow triangles or any other type of empty cycles, the flow will also exhibit a residual \emph{harmonic} component, corresponding to the portion of signal rotating around the holes, as shown in Fig.~\ref{fig:Laplacian_evecs}. 

\subsection{Hodge Laplacians and Hodge Decomposition}
\label{subsec:Hodge}
Having defined the boundary operators and related operations on higher-order signals, we can now introduce the combinatorial Hodge Laplacian (hereafter the Hodge Laplacian). Through this operator, any flow on a simplicial complex can be uniquely split into three components---irrotational, rotational, and harmonic---each living on the edges of the network and representing the portion of the signal responsible for divergence, curl, and harmonic flow, respectively.

\subsubsection{Hodge Laplacian} 
Given an oriented simplicial complex $\mathcal{X}$ of order $K$ with boundary operators $\mathbf{B}_1, \dots, \mathbf{B}_K$, the higher-order Hodge Laplacian $\mathbf{L}_k$ is defined as follows:
\begin{equation}
    \label{eq:HodgeLaplacian}
    \mathbf{L}_k = \bkt \bk + \bkpu \bkput, \quad  0<k<K, 
\end{equation}
whereas the $0$-th and the $K$-order Hodge Laplacians only include the upper or the lower term:
\begin{equation}
     \mathbf{L}_0 = \mathbf{B}_1 \mathbf{B}_1\transpose, \quad \mathbf{L}_K = \mathbf{B}_K\transpose \mathbf{B}_K.   
\end{equation}
For any $0<k<K$, $\mathbf{L}_k$ decomposes naturally into two additive components, which are denominated \emph{lower} (or \emph{down}) Laplacian $\mathbf{L}_{k\downarrow} := \bkt \bk$ and \emph{upper} (or \emph{up}) Laplacian $\mathbf{L}_{k\uparrow} := \bkpu \bkput$, and model different relationships between simplices of the same order. 

In particular, for the case of $1$-simplices, the full Hodge Laplacian $\mathbf{L}_1$ encodes all the neighborhood relationships among edges. The lower Laplacian $\mathbf{L}_{1 \downarrow}$ only captures the adjacencies between edges induced by shared nodes, also referred to as lower adjacencies, as $[\mathbf{L}_{1 \downarrow}]_{ij} \neq 0 $ only if the edges $e_i$ and $e_j$ share a node or $i=j$. Conversely, the upper Laplacian $\lup$ encodes edge adjacencies induced by shared triangles, as $[\mathbf{L}_{1 \uparrow}  ]_{ij} \neq 0 $ only if the edges $e_i$ and $e_j$ are part of a common $2$-simplex in $\mathcal{X}$ or $i=j$.

Note that, similarly to the curl and divergence operators, the Hodge Laplacian considered here is closely related to the continuous Laplace–de Rham operator acting on differential forms, as well as to the Helmholtz decomposition of vector fields \cite{crane2018discrete, ribando2024combinatorial}.

Interestingly, the Hodge Laplacian $\mathbf{L}_1$ fully encodes the structure of a simplicial complex of order $2$. Indeed, due to the inclusion relationships inherent to simplicial complexes, the edges and their participation in triangles implicitly determine the set of nodes in the complex. More generally, for a simplicial complex of order $K$, the $(K-1)$-Hodge Laplacian fully encodes the topological domain and its multi-scale relations: the $(K-1)$-simplices and their assembly into $K$-simplices implicitly define the lower-order simplices as well.

\textbf{Remark: Hodge Laplacian and graph Laplacian} Any graph is a simplicial complex of dimension $1$, and its Hodge Laplacian $\mathbf{L}_0 =  \mathbf{B}_1\mathbf{B}_1\transpose$ corresponds to the standard graph Laplacian $\mathbf{L} = \mathbf{D} - \mathbf{A}$, where $\mathbf{D}$ is the degree matrix, a diagonal matrix with $\left[\mathbf{D}\right]_{i,i}$ corresponding to the number of edges involving node $i$, and $\mathbf{A}$ is the binary adjacency matrix, where $\left[\mathbf{A}\right]_{i,j}$ is $1$ if node $i$ is connected to node $j$, $0$ otherwise.

\subsubsection{Hodge Decomposition}
By using the definition of Hodge Laplacian and the property that $\bk \bkpu = \mathbf{0}$, it is possible to obtain the so-called Hodge decomposition of the signal space, here presented for the edge case, with $E$ number of edges of the simplicial complex. The signal space can indeed be decomposed into the following subspaces:
\begin{equation}
    \label{eq:hodge_decomposition_1}
    \mathbb{R}^{E} = \operatorname{im}(\mathbf{B}_1\transpose) \oplus \ker(\mathbf{L}_1) \oplus \operatorname{im}(\mathbf{B}_2),
\end{equation}
or, equivalently,
\begin{equation}
    \label{eq:hodge_decomposition_2}
    \mathbb{R}^{E} = \operatorname{im}(\mathbf{L}_{1 \downarrow}   ) \oplus \ker(\mathbf{L}_1) \oplus \operatorname{im}(\mathbf{L}_{1 \uparrow}  ),   
\end{equation}
where the direct sum derives from the orthogonality of the spaces. We denote the dimensions of the three spaces as $N_{\downarrow}$, $N_{\text{harm}}$, and $N_{\uparrow}$, respectively\footnote{We refer to the edge space as $\mathbb{R}^E$ since there exists an isomorphism between the space of signals on the $1$-simplices of $\mathcal{X}$ and $\mathbb{R}^E$ when the number of $1$-simplices is finite and the signal is real, as noted in \cite{schaub2021signal}.}.
This decomposition shows that the Hodge Laplacian partitions the space into mutually orthogonal subspaces. In particular, $\operatorname{im}(\mathbf{L}_{1 \downarrow} )$ is the \textit{irrotational} component, corresponding to the subspace capturing signal gradients and contributing to non-zero divergence. Conversely, $\operatorname{im}(\mathbf{L}_{1 \uparrow})$ is known as \textit{rotational} (or solenoidal) component, which contributes to non-zero curl \cite{barbarossa2020topological}. 

While the first two subspaces capture local relationships between adjacent edges, the third one, known as the \emph{harmonic space}, captures global properties of the domain. Indeed, its dimension corresponds to the number of $k$-order holes of the domain, also known as \emph{Betti numbers} and indicated by $\beta_k$ \cite{hatcher2001algebraictopology, nanda2016algebraic, barbarossa2020topological}. 

For instance, Hodge theory implies that the dimension of $\operatorname{ker}(\mathbf{L}_0)$ equals the first Betti number $\beta_0$, which corresponds to the number of connected components of the domain. Similarly, $\beta_1 = \operatorname{dim}(\operatorname{ker}(\mathbf{L}_1))$ counts the number of independent one-dimensional holes, while $\beta_2 = \operatorname{dim}(\operatorname{ker}(\mathbf{L}_2))$ corresponds to the number of cavities, and so on\footnote{More formally, the kernel of the Hodge Laplacian is \emph{isomorphic} to the homology group of the corresponding domain, and its dimension matches the associated Betti number.}.

Importantly, Eq.~\eqref{eq:hodge_decomposition_2} enables the decomposition of any edge signal on $\mathcal{X}$ into the following orthogonal components:
\begin{equation}
    \label{eq:comps}
    \vc{x}^{(1)} = \vc{x}^{(1)}_{\downarrow} + \vc{x}^{(1)}_{\uparrow} + \vc{x}^{(1)}_{\text{harm}},
\end{equation}
where the first two components result from the linear projection of the signal onto the space of $\mathbf{L}_{1 \downarrow} $ and $\mathbf{L}_{1 \uparrow} $, respectively. The first component of Eq.~\eqref{eq:comps} is the irrotational one and is curl-free, whereas the second one is the solenoidal component and is gradient-free. Differently, $\vc{x}^{(1)}_{\text{harm}}$ lies in $\ker(\mathbf{L}_1)$, and correspond to the fraction of signal rotating around holes. 

Alternatively, the same decomposition can be expressed from the perspective of signals of different orders:
\begin{equation}
    \vc{x}^{(1)} = \mathbf{B}_1\transpose \vc{x}^{(0)} + \mathbf{B}_2 \vc{x}^{(2)} + \vc{x}^{(1)}_{\text{harm}},
\end{equation}
with $\vc{x}^{(0)} \in \mathbb{R}^N$ and $\vc{x}^{(2)} \in \mathbb{R}^T$. This formulation clarifies how edge signals can arise as a sum of contributions from signal of adjacent orders, as already discussed in \cite{barbarossa2020topological,schaub2021signal}. In particular, one first contribution to the edge space derives from signals that can be expressed as gradients of nodal scalar potentials $\vc{x}^{(0)}$, which are lifted from the nodal to the edge space by multiplication by $\mathbf{B}_1\transpose$. Conversely, another contribution comes from local circulation patterns supported on 2-simplices of the complex, linearly diffused down to the edge level through multiplication by $\mathbf{B}_2$. Finally, $\vc{x}^{(1)}_{\text{harm}}$ accounts for harmonic components not attributable to either mechanisms that lies in the kernel of these operators. This introduces additional nuances for domain modeling, as discussed in Sec.~\ref{subsec:domain}.

Next, we analyze how spectral properties of the Hodge Laplacians can be leveraged to extend the concept of frequency to simplicial signals, enabling Fourier-style processing.

\subsection{Fourier Goes Topological: Spectral Signal Processing with the Hodge Laplacian}
\label{subsec:Fourier}
A cornerstone of \gls{dsp} is the duality between time and frequency representations of signals, underpinning linear time invariant systems and filters for a broad range of ideas and techniques in \gls{dsp} \cite{roberts1987digital}. However, extending this concept to signals defined on non-Euclidean domains such as graphs is non-trivial, as the notion of a time shift, which underlies this duality, has no direct counterpart on irregular networks \cite{leus2023graph}. One crucial intuition to address this limitation has been to interpret matrices such as the adjacency or Laplacian as \gls{gso} that are aware of the graph structure, analogous to time shifts in \gls{dsp}. For undirected graphs, a \gls{gso} is a symmetric matrix, so its eigendecomposition yields a Fourier-like basis with positive eigenvalues that have a natural interpretation as graph frequencies \cite{ortega2018graph, leus2023graph}. Interestingly, these ideas can be seamlessly extended to \gls{tsp} by defining suitable \gls{tso} \cite{schaub2022signal}. The study of their properties, in particular for the case of the Hodge Laplacian, have fueled the development of a spectral theory for topological signals, as introduced in the seminal work of Barbarossa and colleagues \cite{barbarossa2020topological}, and have enabled the definition of topological linear time-invariant systems and filters \cite{yang2022simplicial}.

Next, we formalize the notions of topological frequency and Fourier transform based on the Hodge Laplacian, and we discuss relevant applications to real signals.

\subsubsection{Spectral Decomposition of the Hodge Laplacian} 
From an algebraic perspective, the Hodge Laplacian is represented by a symmetric matrix. Therefore, similarly to \gls{gsp}, it is possible to introduce its spectral decomposition as follows (here for the case of $\mathbf{L}_1$):
\begin{equation}
    \mathbf{L}_1 = \mathbf{U}_1 \mathbf{\Lambda}_1 \mathbf{U}_1\transpose,
\end{equation}
where $\mathbf{U}_1$ is an orthogonal matrix which has per columns the eigenvectors of $\ma L_1$, and $\mathbf{\Lambda}_1= \operatorname{diag}(\lambda_1, \dots,  \lambda_E)$, whose diagonal entries are the eigenvalues of the matrix. The same holds for any order $k$.

One important consequence of the Hodge decomposition and the orthogonality of the irrotational, rotational, and harmonic spaces is that they can be represented by distinct sets of eigenvectors of the Hodge Laplacian \cite{yang2022simplicial, schaub2021signal, isufi2025topological}. Indeed, $\mathbf{U}_1$ can be obtained by concatenating (and potentially reordering) column-wise the following three matrices,
\begin{equation}
    \mathbf{U} = \left[\mathbf{U}_{1\text{harm}}, \  \mathbf{U}_{1\downarrow}, \ \mathbf{U}_{1\uparrow}\right], 
\end{equation}
where
\begin{itemize}
    \item The columns of $\mathbf{U}_{1\downarrow} \in \mathbb{R}^{E \times N_{\downarrow}}$ are the eigenvectors of $\ma L_{1\downarrow}$ with eigenvalues $>0$, spanning the irrotational space;
    \item The columns of $\mathbf{U}_{1\uparrow} \in \mathbb{R}^{E \times N_{\uparrow}}$ are the eigenvectors of $\ma L_{1\uparrow}$ with eigenvalues $>0$, spanning the solenoidal space;
    \item The columns of $\mathbf{U}_{1\text{harm}} \in \mathbb{R}^{E \times N_{\text{harm}}}$ are the eigenvectors of $\ma L_1$ associated with zero eigenvalues, lying in $\operatorname{ker}(\ma L_1)$, and they span the harmonic component of the edge space.
\end{itemize}

Due to the symmetry of $\mathbf{U}_1$, its eigenvalues are all real and positive; therefore, similarly to \gls{gsp}, they can be interpreted as frequencies. Broadly, these frequencies quantify the degree of variation of each associated eigenvector over the domain, with lower frequencies being associated to smoother variations over the edge set \cite{ortega2018graph,yang2022simplicial,barbarossa2020topological2}. 

Notably, as the three sets of eigenvectors capture distinct types of patterns in the domain, their frequencies are also associated to different types of variations \cite{yang2022simplicial, barbarossa2020topological, schaub2021signal}, as illustrated in Fig.~\ref{fig:Laplacian_evecs}. In particular, eigenvectors in $\mathbf{U}_{1\downarrow}$ with small non-zero eigenvalues correspond to gradients oscillating slowly across the edges following smooth, irrotational patterns. Higher eigenvalues correspond to higher-frequency gradient components, encoding more rapid variations along the edges. Conversely, eigenvectors in $\mathbf{U}_{1\uparrow}$ capture rotational patterns, with small eigenvalues corresponding to slow, large-scale cycles around the domain, and larger eigenvalues capturing ``faster", more localized rotations.

One way to give meaning to frequency ordering and understand the relationship between eigenvalues and different types of variations is through their connection to quadratic forms. For example, for any gradient frequency $\lambda$ associated with a lower eigenvector $\vc{u}_{\downarrow}$, the following holds \cite{yang2022simplicial, isufi2025topological, schaub2021signal}:
\begin{equation}
    \label{eq:frequenciesenergy}
    \begin{aligned}
    \lambda
        & = \vc{u}_{\downarrow}\transpose \mathbf{L}_{1\downarrow}\vc{u}_{\downarrow} = \lVert \mathbf{B}_1 \vc{u}_{\downarrow} \rVert_2^2\\
        &= \lVert \mathbf{B}_1 \vc{u}_{\downarrow} \rVert_2^2
           + \lVert \mathbf{B}_2\transpose \vc{u}_{\downarrow} \rVert_2^2 \\
        &= \lVert \mathbf{B}_1 \vc{u}_{\downarrow}
           + \mathbf{B}_2\transpose \vc{u}_{\downarrow} \rVert_2^2 \\
        &= \vc{u}_{\downarrow}\transpose \mathbf{L}_{1}\vc{u}_{\downarrow},
    \end{aligned}
\end{equation}
where the last term represents the overall quadratic variation with respect to the Hodge Laplacian. Note that the second equality directly follows from the definition of $\mathbf{L}_{1\downarrow}$, while the third one is justified by the fact that $\lVert \mathbf{B}_2\transpose \vc{u}_{\downarrow} \rVert_2^2=0$, a consequence of the orthogonality of different sets of eigenvectors. Finally, the last two follow from the orthogonality of the different subspaces induced by the Hodge decomposition and the definition of $\mathbf{L}_1$. Note that $\lVert \mathbf{B}_1 \vc{u}_{\downarrow} \rVert_2^2
= \lVert \operatorname{div}(\vc{u}_{\downarrow}) \rVert_2^2$ is the squared $l_2$-norm of the divergence of the eigenvector $\vc{u}_{\downarrow}$, highlighting that the frequency content of an eigenvector in the irrotational space is a function of its divergence. Similarly, each eigenvalue of $\mathbf{L}_{1\uparrow}$ corresponds to the squared $l_2$-norm of the curl of its associated eigenvector. Finally, the harmonic eigenvectors, lying in the kernel of $\mathbf{L}_1, \mathbf{L}_{1\downarrow}$ and $\mathbf{L}_{1\uparrow}$, correspond to zero-frequency components. Most interestingly, due to the deep connection between the Hodge Laplacian and the topological properties of the space discussed above, the harmonic eigenbasis has the same dimensionality as the holes in the domain, with each eigenvector encoding a rotation around a specific hole, as illustrated in Fig.~\ref{fig:Laplacian_evecs}. Further considerations on the spectral properties of the Hodge Laplacian eigenvalues can be found in \cite{grande2024disentangling}.

\textbf{Remark: limitations of the frequency analogy} As in \gls{gsp}, the analogy with the classical notion of frequency is limited: eigenvectors with higher topological frequencies can be highly localized on a subset of edges depending on the network structure, so the eigenvalue ordering does not always provide an intuitive interpretation \cite{ortega2018graph, leus2023graph}.

The spectral perspective on topological operators enables the simultaneous analysis of signals according to both type and related frequency content. As in \gls{dsp} and \gls{gsp}, the key operation for assessing a signal’s frequency content is the Fourier transform, now reformulated for \gls{tsp}.

\subsubsection{Topological Fourier Transform}
The \gls{tft} of a signal $\vc{x}^{(k)}$ is defined as its projection onto the space of the eigenvectors $\mathbf{U}_k$ of the Hodge Laplacian $\ma L_k $ as follows:
\begin{equation}
    \tilde{\vc{x}}^{(k)}= \mathbf{U}_k\transpose\vc{x}^{(k)},
\end{equation}
where the $i$-th entry $\tilde{x}^{(k)}_i$ quantifies how strongly the signal aligns with the $i$-th frequency; in particular, a high value indicates that a large amount of the signal energy is concentrated at that frequency. Following the orthogonality of the components from different subspaces, the vector of Fourier coefficients can be obtained by concatenating the coefficients from the different subspaces 
$\tilde{\vc{x}}^{(k)} =\left[\tilde{\vc{x}}^{(k)}_{\downarrow},  \tilde{\vc{x}}^{(k)}_{\uparrow}, \  \tilde{\vc{x}}^{(k)}_{\text{harm}}\right]$,
where the first two sets of coefficients are the ones associated to the non-zero frequencies of the irrotational and solenoidal space, respectively, whereas the last ones are associated to the harmonic frequencies \cite{barbarossa2020topological}.

Notably, the generalization of the Fourier transform to topological domains provides the theoretical foundation for extending important filtering concepts and properties to \gls{tsp}, as discussed in the following paragraph.

\begin{insightbox}{Trajectory Embedding Using Hodge Eigenvectors}
{
    An increasingly explored application of the Hodge Laplacian spectral decomposition involves the classification and clustering of trajectories. In this setting, a trajectory on a discretized geographical domain can be represented as an edge signal, where each segment of the trajectory is encoded by a signal on the corresponding edge of the underlying discretized domain, as done in \cite{schaub2020random}.
    
    Projecting the trajectory onto the eigenvectors of the Hodge Laplacian yields its topological Fourier coefficients, computed via inner products with the corresponding eigenvectors. Representing the signal through a subset of these coefficients, such as the ones linked to eigenvectors with lowest frequencies, provides a compact spectral representation suitable for downstream tasks such as classification and clustering. This representation is referred to as \emph{embedding} of the trajectory (into the corresponding Fourier subspace). 

    For instance, examining the harmonic embedding of a trajectory provides insight into its position relative to the domain holes, owing to the one-to-one correspondence between holes and harmonic eigenvectors. This approach has been leveraged in several works, both on synthetic \cite{schaub2021signal} and real data, for a range of tasks including denoising street flows \cite{schaub2018flow}, predicting street mobility \cite{ghosh2018topological}, clustering patient flows across healthcare providers \cite{gebhart2021go} or trajectories of ocean currents \cite{schaub2020random}, and tracking cell differentiation \cite{cheng2025phlower}. 
}
\end{insightbox}

\subsubsection{Simplicial Convolutional Filters} 
Filtering operations lie at the heart of signal processing, with convolutional filters representing a specific class that exploits properties in both the signal and frequency domains, forming the basis of linear time-invariant systems.

Simplicial convolutional filters have recently been introduced by Yang and colleagues as polynomial functions of the Hodge Laplacian \cite{yang2022simplicial}, in analogy with \gls{gsp} \cite{ortega2018graph}. In particular, a convolutional filter acting on the edges of a simplicial complex $\mathcal{X}$ with associated Hodge Laplacian $\mathbf{L}_1$ can be written as a polynomial function of the components of $\mathbf{L}_1$ as follows:
\begin{equation}
  h(\mathbf{L}_1) = h_{\text{harm}}\mathbf{I} + \sum_{m=1}^{M_\downarrow} \alpha_m \mathbf{L}_{1 \downarrow}   ^m + \sum_{n=1}^{M_\uparrow} \beta_n \mathbf{L}_{1 \uparrow}^n,
\end{equation}
\noindent where $\mathbf{I} \in \mathbb{R}^{E \times E} $ is the identity matrix, $M_\downarrow$ and $M_\uparrow$ are the polynomial degrees, $\alpha_m, \beta_n \in \mathbb{R}$ are scalar coefficients of the polynomials in $\mathbf{L}_{1 \downarrow}$ and $\mathbf{L}_{1 \uparrow}$, respectively. Crucially, this local representation highlights how the filter respects the topology of the simplicial complex: after a single pass, each edge updates its state based on information from adjacent edges in both the upper and lower neighborhoods, along with a constant component from the harmonic space. Interestingly, the degree of the upper and lower components of the polynomial determines the size of the neighborhood considered. For example, a convolutional filter with lower degree $1$  (i.e., $M_{\downarrow}=1$) updates each edge based on the weighted values of its immediate topological lower neighbors, with the weights being determined by the filter coefficients, and the local neighborhood by the nonzero entries of the Hodge Laplacian associated to that edge. For instance, if the polynomial has degree 2, each edge is influenced not only by its immediate lower neighbors, but also by the immediate lower neighbors of those neighbors. More generally, for a polynomial of degree $k$, each edge is affected by its so-called $k$-hop neighborhood, capturing increasingly larger portions of the simplicial complex. As noted in \cite{yang2022simplicial}, a consequence of the Cayley--Hamilton theorem is that each convolutional filter admits a polynomial representation whose degree is bounded by the number of distinct irrotational and solenoidal frequencies. In particular, $M_\downarrow \leq D_\downarrow, M_\uparrow \leq D_\uparrow,$ where $D_\downarrow$ and $D_\uparrow$ denote the numbers of distinct gradient and curl frequencies, respectively. 

A significant advantage of the above definition is that contributions from the lower and upper adjacent simplices are treated separately and then combined, while the harmonic component provides a uniform contribution corresponding to the zero-order term. Therefore, the filter can be implemented as a sequence of local shift-and-sum operations. Furthermore, from a practical perspective, this formulation naturally enables a distributed filter implementation.

Most importantly, in analogy with \gls{dsp} and \gls{gsp}, convolutional filtering in the simplicial domain corresponds to pointwise multiplication in the Fourier domain, yielding a compact and flexible spectral representation of the filter. Below, we report the main results from \cite{yang2022simplicial}. In particular, any filter $\ma H$ which is a polynomial of the Hodge Laplacian is jointly diagonalizable with it, indicating that admits the same Fourier basis. By diagonalizing $\ma H$ via $\ma U_1 = [\ma U_{1\text{harm}} \ \ma U_{1\downarrow} \ \ma U_{1\uparrow}]$, we obtain the frequency representation of $\ma H_1$ as follows (\cite[Sec.~C]{yang2022simplicial}:
\begin{equation}
    \tilde{\ma H} = \ma U_1\transpose \ma H \ma U_1 = \mathrm{blkdiag}(\tilde{\ma H}_{\text{harm}}, \tilde{\ma H}_\downarrow, \tilde{\ma H}_\uparrow),
\end{equation}
where $\tilde{\ma H}_{\text{harm}}, \tilde{\ma H}_\downarrow, \tilde{\ma H}_\uparrow$ are diagonal matrices. Equivalently, the spectral response at frequency $\lambda$ is given by
\begin{equation}
    \tilde{\ma H}(\lambda) =
    \begin{cases}
    h_{\text{harm}}, & \lambda =0, \\
    h_{\text{harm}} + \sum_{m=1}^{M_\downarrow} \alpha_{m} \lambda^{m}, & \lambda \in \operatorname{diag}(\ma{\Lambda}_\downarrow), \\
    h_{\text{harm}} + \sum_{n=1}^{M_\uparrow} \beta_{n} \lambda^{n}, & \lambda \in \operatorname{diag}(\ma{\Lambda}_\uparrow),
    \end{cases}
\end{equation}
where $\operatorname{diag}(\ma{\Lambda}_\downarrow)$ denotes the set of gradient frequencies, and $\operatorname{diag}(\ma{\Lambda}_\uparrow)$ the set of curl ones. A direct consequence of this property is that the convolution theorem can be extended to simplicial complexes, whereby convolution in the original domain corresponds to multiplication in the Fourier domain. Indeed,
\begin{equation}
\ma H \vc x = (\ma U_1 \tilde{\ma H} \ma U_1 \transpose)(\ma U_1 \tilde{\vc x} \ma U_1 \transpose)
= \ma U_1 \tilde{\ma H} \tilde{\vc x} \ma U_1\transpose.
\end{equation}
The block-diagonal structure of $\tilde{\ma H}$ implies that filtering reduces to pointwise multiplication of each spectral component of the signal in the Fourier domain, $\tilde{\vc x}$, by the corresponding frequency response of the filter, which can be followed by an inverse Fourier transform to map the signal back to the original domain.
By leveraging the spectral decomposition of the Hodge Laplacian, it is therefore possible to selectively amplify or attenuate signal components according to both their type and frequency. For instance, low-frequency irrotational components can be retained to capture smooth, gradient-like trends across the network, whereas low-frequency rotational components can be emphasized to reveal macroscopic cyclic or loop-like patterns  (see \cite{yang2022simplicial} for further details).

\begin{insightbox}{Applications of Simplicial Filters}
    All fundamental signal processing tasks on real data, such as component extraction and denoising, can be formulated in terms of filtering operations. In \cite{yang2022simplicial}, the authors show that several of these tasks can be efficiently implemented via convolutional filtering. For instance, they show how the relative importance of streets in London can be assessed using an edge-based variant of the PageRank algorithm (see \cite{schaub2020random} for a description), which can be expressed as an appropriate polynomial filter of the Hodge Laplacian.

    Beyond specific tasks, filtering is also motivated by computational considerations. In the same work, it is shown that polynomial filters enable the extraction of gradient- and curl-related components of traffic signals on the Chicago road network. While the same decomposition could in principle be obtained by projecting onto the corresponding Hodge Laplacian components, polynomial filters avoid matrix inversion, leading to a more efficient and scalable implementation.
    
    Moreover, topological filters have emerged as a key tool for building neural networks that incorporate the topological structure of the underlying domain. In this context, simplicial convolutional filters have become central to the construction of a broad range of \gls{tnn} architectures, first introduced in \cite{yang2023convolutional, yang2022simplicial2}. In most of these models, each layer effectively performs a filtering operation on the data defined via a polynomial filter of a \gls{tso} (often the Hodge Laplacian), where the filter parameters are learnable weights of the network. This induces a form of structure-aware processing, in which the model is encouraged to respect the biases introduced by the domain topology \cite{isufi2025topological, papamarkou2024position, hajij2022topological}.
\end{insightbox}

\subsection{Links Between \gls{gsp} and \gls{tsp}}
\label{subsec:gsp_tsp}
As previously mentioned, any graph is also a simplicial complex. This implies that all concepts and operations defined in the \gls{gsp} framework are in fact special cases of their more general \gls{tsp} counterparts. To elaborate on this link, the following paragraphs discuss some analogies and connections between these two frameworks.

\subsubsection{Graph Laplacian and Edge Laplacian} 
As a simplicial complex of order $1$, any graph with $N$ nodes and $E$ edges can be fully characterized by its zeroth-order Laplacian
$\mathbf{L}_0 = \mathbf{B}_1 \mathbf{B}_1\transpose \in \mathbb{R}^{N \times N},$ which consists solely of the upper Laplacian component $\mathbf{L}_{0 \uparrow}$. This operator coincides with the combinatorial graph Laplacian $\mathbf{L}$. Interestingly, the same network can be described from an edge-centric perspective by encoding adjacency relations between edges via the lower component of the $1$-Hodge Laplacian, $\mathbf{L}_{1 \downarrow} = \mathbf{B}_1\transpose \mathbf{B}_1 \in \mathbb{R}^{E \times E}$, which in this context is also referred to as \emph{edge Laplacian} \cite{schaub2021signal}. Since $\mathbf{L}_0$ and $\mathbf{L}_{1 \downarrow}$ are constructed as $\mathbf{B}_1 \mathbf{B}_1\transpose$ and $\mathbf{B}_1\transpose \mathbf{B}_1$, respectively, they share the same non-zero eigenvalues, although their kernels have different dimensions. Therefore, each nonzero eigenvalue corresponds to a pair of eigenvectors, one in the node space and one in the edge space \cite{yang2022simplicial}. 

In particular, for a given pair of eigenvectors $\vc{u}^{(0)}_{j}$ and $\vc{u}^{(1)}_{j}$ associated with the same non-zero eigenvalue $\lambda_j$, the corresponding node eigenvector can be obtained from the edge eigenvector via the boundary operator as shown below:
\begin{equation}
    \vc{u}^{(0)}_{j} = \frac{1}{\sqrt{\lambda_j}} \vc{B}_1 \vc{u}^{(1)}_{j}.
\end{equation}

Indeed, this follows directly from the definition of eigenvectors of $\ma{L}_0$:
\begin{equation}
\begin{aligned}
    \ma{L}_0 (\vc{B}_1 \vc{u}^{(1)}_j)
    &= \vc{B}_1 \vc{B}_1^\top \vc{B}_1 \vc{u}^{(1)}_j
    = \vc{B}_1 \ma L_{1\downarrow} \vc{u}^{(1)}_j\\
    &= \vc{B}_1 (\lambda_j \vc{u}^{(1)}_j)
    = \lambda_j (\vc{B}_1 \vc{u}^{(1)}_j).
\end{aligned}
\end{equation}

The normalization follows from the fact that $\|\vc{B}_1 \vc{u}^{(1)}_j\|_2^2 = \lambda_j$, as shown in Eq.~\eqref{eq:frequenciesenergy}. Vice versa, the nodal eigenvector can be mapped to the edge one by multiplication by $\ma B_1\transpose$ up to scaling. This link clarifies how the nodal and edge eigenvectors yield related patterns, with $\mathbf{L}_{1 \downarrow}$ offering an edge-centric, flow-based view of the structure captured at the nodal level by $\mathbf{L}_0$, and vice versa. An example of this concept on a brain imaging graph is shown in Fig.~\ref{fig:brain_eigenvectors}. Notice, however, that this representation is completely blind to the rotational and harmonic information, that is instead encoded in the eigenvectors of $\ma L_{1\uparrow}$ and $\ma L_{1 \text{harm}}$. Indeed, for any signal $\vc x^{(1)}$, its rotational and harmonic components lie in the kernel of the edge Laplacian.

\subsubsection{Line Graph}
A widely used approach to process edge signals with \gls{gsp} tools relies on the construction of the line graph \cite{betzel2023living}. Given an original graph, its line graph is obtained by treating edges as new graph nodes, with the adjacency relationships between edges in the original graph defining the new edges in the line graph. Edge signals are then treated as node signals on the resulting line graph \cite{betzel2023living}.

While this approach is suitable for signals without a natural orientation, it presents several important limitations when dealing with flow-like data. In particular, the line graph captures only lower adjacencies between edges through shared nodes, while neglecting upper adjacencies through incident triangles. As a result, higher-order topological information cannot be extracted from the signal. Moreover, by collapsing the edge domain into a nodal one, edge orientation is not preserved, and the sign of each signal entry loses its initial interpretation. This, in turn, leads to degraded performance in tasks such as denoising, as compared to $\ma L_{1}$. A more detailed discussion of this aspect, together with useful illustrative examples, is provided in \cite[Sec.~4]{schaub2021signal}, which we only briefly summarize here. The key intuition highlighted by the authors is that this modeling choice strongly affects the notion of smoothness implicitly leveraged by the filters. In particular, smoothing based on the line graph implicitly assumes that adjacent edges should carry similar signal values. However, for flow-like data, the individual flows on neighboring edges can exhibit substantial variability, even when the overall flow entering and exiting each node is preserved, which motivates the use of the Hodge Laplacian.

%% file: Main_text/Sections/2_Domain_and_Edge_Signals.tex
\section{Challenges in Applying Topological Signal Processing: Domains and Signals}
\label{sec:Domain_and_Edge_Signals} 

\noindent We now address the challenges of applying \gls{tsp} to real data when at least one of its key ingredients---the domain or the signal---is not readily available and has to be derived from measures of lower order. First, we briefly review the main approaches used to model networks with simplicial complexes when higher-order links or relationships are not explicitly encoded in the dataset. Second, we discuss how higher-order signals can be computed or inferred from lower-order ones, and we outline the properties that make them most suitable for \gls{tsp}-based analysis.  As approaches to infer the domain have been more extensively discussed in the literature, we place greater emphasis here on the challenges related to the signal. We focus in particular on signals of order 1, expressed on the edge of a simplicial complex. 
As an illustrative example, we show how an edge signal can be derived from nodal time series capturing temporally lagged dynamics, and we derive some of its basic theoretical properties, which are leveraged in the case study presented in Sec.~\ref{sec:Application}.

\subsection{High-Order Domains}
\label{subsec:domain}
Mathematically, inferring a higher-order domain from data amounts to defining the entries of \gls{tso}, such as the Hodge Laplacian, under assumptions on how the data relate to the latent domain. To this aim, we adopt a signal processing-driven perspective and categorize existing approaches based on the \gls{tsp} components used to estimate the simplicial complex. From this viewpoint, one class of methods focuses on estimating the simplicial complex directly from the data defining the support, such as a point cloud of observations or a (partially observed) graph. A second class of approaches leverages instead the observed signals and assumptions on their regularity over the underlying domain to infer the latent network higher-order structure. Note that this categorization is not exhaustive, and the distinction between the two classes may sometimes be blurred, as certain methods can combine elements of both. Rather than reviewing every approach in detail, our exposition aims to provide a road map to help the reader navigate the literature and understand the rationale behind different modeling choices.

\subsubsection{Deriving Simplicial Complexes from Domain Data}
In this setting, the goal is to construct a simplicial complex directly from the data defining the domain, without considering the signal. This is a broad and fundamental problem in \gls{tda} \cite{carlsson2021topological, wasserman2018topological}.

\paragraph{Starting from Point Clouds: Geometry-Based Constructions}
\label{par:domain_geometry}
When data are provided as a point cloud; i.e., a set of points in a metric space, one viable approach is to build a simplicial complex relying on geometric constructions based on distances between points, such as the Vietoris–Rips or \v{C}ech complexes \cite{carlsson2021topological, wasserman2018topological}. Furthermore, if the ultimate goal of the analysis is to infer topological invariants---such as holes---from the point cloud, these constructions can be used to compute \emph{persistent homology}, a cornerstone of \gls{tda}. For more details, see \cite{carlsson2021topological, wasserman2018topological} or any other standard text on \gls{tda}. Other geometric constructions, such as triangulations induced by the point cloud (e.g., Delaunay triangulations), can also be employed to obtain meshes or other discretizations of the space, as illustrated in Fig.~\ref{fig:Laplacian_evecs}. Some examples of this approach, applied to generate simplicial complexes from geographic data, are provided in \cite{schaub2020random, yang2023hodge, chen2021helmholtzian}.

\paragraph{Starting from Networks: Clique Complexes and Threshold-Based Approaches}
If the initial data already includes a partially observed network structure, for instance a graph, one can build a valid simplicial complex of order $2$ by ``filling'' all the possible hollow $2$-simplices composed by triplets of graph edges (i.e., the $3$-cliques). This is also the approach followed in the case study presented in Sec.~\ref{sec:Application}. Other approaches can be threshold-based. Consider, for instance, a dataset with $N$ nodes and a graph structure encoded in a weighted matrix $A \in \mathbb{R}^{N\times N}$, where the entries of A measure the strength of the pairwise connections. One can decide to add the $2$-simplex $\{v_i, v_j, v_k\}$ when the $(i,j)$, $(j,k)$, and $(k,i)$-th entries of $A$ all exceed a specified threshold. Similarly, if a measures of triadic relationship can be computed directly from the data, one can decide to leverage it and retain only the $k$ $2$-simplices associated with the highest values of that measure, as done in \cite{bispo2025learning}. In all these cases, the threshold can be determined \emph{a priori} from problem-specific knowledge, in a data-driven manner, or to satisfy computational constraints on the size of the simplicial complex.

\subsubsection{Deriving Simplicial Complexes Exploiting Signal Properties}
In this setting, the goal is to infer a partially unobserved network topology from observed signals living on it. This class of methods relies on the assumption that measured signals arise from a process evolving on the network and are therefore intrinsically linked to its topology; for instance, they are smooth over it. This perspective can be regarded as a generalization of the graph-based network inference framework surveyed in \cite{mateos2019connecting}, which offers a comprehensive treatment of the topic. In particular, in the \gls{tsp} case, only a partial observation of the hierarchical structure may be available—e.g., only a graph, a subgraph, or a subset of simplices. To this aim, several data-driven methods infer the localization of $2$-simplices based on criteria such as the minimization of signal energy in specific subspaces \cite{barbarossa2020topological} or probabilistic modeling via Bayesian inference \cite{gurugubelli2024simplicial}. We refer the reader to \cite[Ch.~4.2]{isufi2025topological} for a detailed overview of these methods in the context of \gls{tsp}. 

Beyond the approaches reviewed therein, we highlight an emerging and still rather unexplored direction: extending network inference methods based on graph stationarity (see the related section in \cite{mateos2019connecting}) to the \gls{tsp} setting, leveraging the notion of weak stationarity for signals on simplicial complexes introduced in a recent preprint \cite{navarro2026stationarity}. 

\begin{insightbox}{\textbf{What You Model Shapes What You Can See}}
    One approach that can guide domain modeling is to consider how design choices affect downstream analysis, in particular by identifying higher-order structures that induce a meaningful organization of signal dynamics. Indeed, although the domain is often constructed from pairwise dependencies, introducing higher-order structures such as simplices allows for more nuanced signal representation and processing. 
    
    As an example, including a $2$-simplex allows one to analyze flow circulations around it using richer signal processing tools, such as penalizing non-smooth signals or enhancing specific topological frequencies \cite{schaub2021signal}. On the other hand, if the triangle is not included, such circulations contribute to the harmonic components and lie entirely in the kernel of the Hodge Laplacian. In the harmonic subspace, spectral filtering loses expressive power, in the sense that Laplacian-based frequency resolution is lost, since all harmonic components belong to the zero-eigenvalue eigenspace.
    
    Therefore, starting from a graph with no prior knowledge of the location or number of $2$-simplices, one possible strategy is to include all of them (i.e., construct the corresponding $3$-clique complex), so that all potential circulations are retained and not absorbed into the kernel. By contrast, a more parsimonious strategy consists in selectively including $2$-simplices that capture relevant and smooth circulation patterns in the signal, while omitting those that are less relevant to explain its dynamics \cite{schaub2021signal}. For instance, in a recent preprint applying \gls{tsp} to brain neuroimaging \cite{bispo2026multimodal}, the authors estimate the higher-order elements of a simplicial complex on a brain graph 
    by imposing spectral sparsity on the solenoidal and harmonic components of the observed signals. This perspective aligns with the goal of many signal processing approaches of inferring a topology that guarantees desirable signal properties---such as smoothness over the domain---while remaining sufficiently sparse to admit a parsimonious signal representation (see also \cite{hoppe2024representing}).
\end{insightbox}

A common caveat across many of the methods discussed is that they require the number of higher-order simplices in the complex to be specified a priori, leaving this choice to cross-validation or domain knowledge. This represents an active and evolving area of research.

\subsection{High-Order Signals}
\label{subsec:signals}
As discussed in Sec.~\ref{subsec:signals}, \gls{tsp} thrives in handling signals supported on higher-order domains. However, in most applications, measurements are collected at pointwise locations and hence fall into the category of nodal signals. Nevertheless, deriving higher-order signals from lower-order measurements can be of interest in several scenarios. For example, if an underlying process is more naturally described on the edges of a graph, nodal measurements can be viewed as partial and indirect observations of a latent edge signal, whose inference may provide a more faithful representation of the process. Alternatively, higher-order signals may be derived from lower-order measurements if they provide more informative or useful features, irrespective of the modeling assumptions. However, it is not always immediate to determine when such constructions are appropriate or how they should be carried out in practice.
The goal of this subsection is therefore to dissect the characteristics that make an edge signal suitable to \gls{tsp} analysis, especially when derived from nodal measurements. Importantly, in agreement with what was discussed in previous sections, we mostly examine signals living on oriented edges, as those that mostly benefit from the \gls{tsp} framework. While we concentrate on the specific case of edge signals, the considerations in this chapter can be seamlessly generalized to signals of any order. 

Following a similar logic to the previous subsection, we first introduce a classification of edge signals to help structure the variety of examples found in the literature. Then, as a use case, we introduce an example of an edge signal capturing lead-lag dynamics between node pairs, and we discuss its basic properties. Until the end of this section, we assume a simplicial complex $\mathcal{X}$ of order $2$ with $N, E$, and $T$ cardinality of the nodes, edges, and triangles in the complex, respectively. Moreover, to simplify the notation, in this section only we refer to nodal signals as $\vc{x} \in \mathbb{R}^{N}$ and to edge signals as $\vc{e} \in \mathbb{R}^{E}$.

\subsubsection{What it Means to Be an Edge}
\label{subsec:desiderata}

\paragraph{Different Constructions of Edge Signals}
Following the objective of this paper, we adopt a constructive and application-oriented perspective and classify edge signals according to the type of data from which they are derived: direct edge measurements, nodal measurements, or measurements on simplices of dimension higher than one.

\textbf{Measurements on the edges of a network.} These measurements include, for instance, all types of physical and abstract flows, such as the ones in water, traffic, or telecommunication networks, financial transactions, or message exchanged in a social network. In general, any measurement capturing an oriented pairwise relation between nodes can be naturally represented as an edge signal in the \gls{tsp} framework. For instance, this category includes pairwise rankings of nodes, where a positive signal from node $i$ to $j$ may indicate that node $i$ ranks higher than node $j$, such as the one used in the context of the simplicial PageRank algorithm \cite{schaub2020random}. Another example involves trajectory data: trajectories in a space can be modeled as edge signals by discretizing the space, aggregating data over elements of the discretization---such as triangles, hexagons, etc.---and then defining an (oriented) indicator function supported on the edges of the resulting mesh traversed by the trajectory \cite{schaub2020random, schaub2021signal}. Moreover, this broad category also includes vector fields defined on meshes, which can be represented as signals on the edges of the mesh by projecting the vector at each vertex onto the incident edges of the triangulation, as also discussed in \cite[Sec.~5]{isufi2025topological} and employed in practice in \cite{yang2023hodge, chen2021helmholtzian, barbarossa2020topological}. The generated edge flow can then be processed with \gls{tsp} approaches and then back-projected to vector field. See, for instance, \cite[Sec.~6]{barbarossa2020topological} for the steps required to move between the two representations in an example application.

\textbf{Signal derived from nodal measurements} We are now interested in adopting a \emph{bottom-up} approach and deriving an edge signal from a nodal one. This category includes signals derived from nodal measurements using external priors, where additional knowledge is incorporated to infer edge-level representations. One example is the use of physical or geometrical priors. For instance, by applying Hagen–Poiseuille’s law, it is possible to derive edge-level flow estimates from the difference between nodal pressure data. Another option is to derive them from nodal measures exploiting other dimensions of the signal. In this scenario, edge-level signals are reconstructed or estimated from the joint dynamics over time (or along any other additional dimension) of the nodal signals themselves. For example, lagged dependencies between nodes can themselves be treated as a signal flowing along the edges connecting the nodes, possibly reflecting some form of communication patterns. Notably, the edge signal introduced in the next paragraphs falls within this class. 

\textbf{Signal derived from measurements on higher-order simplices} Edge signals can also be obtained in a \emph{top-down} manner, by down-projecting or manipulating higher-order signals. Given its limited relevance to current applications, we do not elaborate further on this scenario.

\paragraph{Go With the Flow:\\ Numerical Desideratum from a \gls{tsp} Perspective}
From a numerical perspective, an edge signal is said to be consistent with the orientation of the simplex if it changes sign when the indices are exchanged; that is, given an edge signal $e_{i,j}$ defined between nodes $i$ and $j$, it satisfies $e_{i,j} = -e_{j,i}$. This property is known as antisymmetry, and the matrix $\ma E$ representing such a signal is skew-symmetric, meaning that its transpose is equal to its negative: $\vc E \transpose= - \vc E$. From a computational perspective, this properties ensures sign consistency when working with boundary matrices and derived operators (Sec.~\ref{subsec:hodge_boundary}). Intuitively, this means that a flow from node $i$ to node $j$ must carry the opposite sign of a flow from $j$ to $i$, since the two represent the same quantity viewed from opposite reference points.

\paragraph{Candidate Edge Signals}
We briefly review two candidate edge signals derivable from nodal time series, and discuss to which extent they are suitable for a \gls{tsp}-based analysis.

\textbf{Instantaneous products of nodal signals} Signals living on the edges of a network can be obtained as pairwise products of nodal time series. They are frequently used, for instance, in brain imaging literature \cite{betzel2023living}, and interpreted as instantaneous co-fluctuations reflecting coordinated activity between brain regions. Although appealing due to their simplicity and intuitive interpretation, this signal does not satisfy the antisymmetry property; indeed the following holds:
\begin{equation}
    \label{eq:edge_signal_1}
    e_{i,j}[t]=x_i[t]\cdot x_j[t]=e_{j,i}[t].
\end{equation}
This quantity represents a joint, symmetrical (hence non-oriented) relationship between nodes, which is more naturally interpreted as a scalar, density-like measure rather than a flow, with the negative sign encoding disagreement between nodal activity. Therefore, these type of signals are more suited for analysis using the line graph framework, where the assumption of orientation is not required nor leveraged, as pointed out in Sec.~\ref{subsec:gsp_tsp} (see also \cite{schaub2021signal, schaub2018flow}). The same considerations apply to any edge signal obtained as a function $f$ that is symmetric in its arguments, i.e., $e_{i,j}=f(x_i, x_j) = f(x_j, x_i)=e_{j,i}$, such as $\cos\big(\phi(x_i[t]) - \phi(x_j[t])\big)$, where $\phi$ denotes the instantaneous phase, used for instance in \cite{bispo2026multimodal, Santoro2025EdgeLaplacians}. 

\textbf{Instantaneous differences between nodal signals}
The edge signal defined as 
\begin{equation*}
    e_{i,j}[t] = x_i[t] - x_j[t],
\end{equation*} 
meets the antisymmetry requirement by definition, and thus belongs to the class of signals that are particularly amenable to analysis with \gls{tsp}. However, edge signals defined as nodal differences reside entirely in the curl-free (irrotational) component of the edge space. This is a direct consequence of their representation as \emph{gradients} of  nodal signals. Indeed, the signal can be computed as $ \vc{e}=\mathbf{B}_1\transpose \vc x$. Consequently, it holds that $\mathbf{B}_2\transpose \vc{e} =\mathbf{B}_2\transpose \mathbf{B}_1\transpose \vc{x}=(\mathbf{B}_1 \mathbf{B}_2)\transpose \vc{x}= \vc{0}$, so that $\vc e$ has zero curl. Equivalently, the sum of edge signals around any closed path---including triangular ones around $2$-simplices---vanishes identically. As a result, this signal provides an edge-based representation of the original nodal dynamics in terms of gradients, which cannot exhibit circulations around $2$-simplices by construction. In any case, it can be interpreted as a flow and is therefore amenable to analysis using the lower Laplacian $\mathbf{L}_{1\downarrow}$.

\subsection{A Guiding Example:\\ Edge Signal Capturing Lead-Lag Nodal Dynamics}
\label{subsec:proposed_edge_signal}
In light of the previous considerations, we propose here a simple definition of node-derived edge signal suitable for \gls{tsp}-based analysis that leverages the temporal structure of the original time series.

\subsubsection{Definition}
\label{subsec:definition}
Given the signals $\vc{x}_i$ and $\vc{x}_j$ defined over time and associated with two connected nodes $i$ and $j$ of a simplicial complex $\mathcal{X}$, the proposed edge signal is defined as
\begin{equation}
    \label{eq:edge_signal_main}
    e_{i,j}[t] = x_{i}[t-1] x_{j}[t] - x_{j}[t-1] x_{i}[t], \quad t \geq 1.
\end{equation}
This expression is inherently antisymmetric in the node indices and explicitly couples the activity of nodes $i$ and $j$ across two consecutive time points. Equivalently, it can be reformulated expliciting the role of the temporal increments:
\begin{equation}
    \label{eq:edge_signal_lag}
  e_{i,j}[t] = x_i[t-1] \Delta x_j[t] - x_j[t-1] \Delta x_i[t],  \quad t \geq 1
\end{equation}
where $\Delta x_j[t] = x_j[t] - x_j[t-1]. $ Interestingly this signal definition can be highlighted from different perspectives, including as a measure of lead-lag dynamics.

\subsubsection{Interpretation as a Determinant}
\begin{figure*}[htbp]
  \centering
  \includegraphics[width=.85\linewidth]{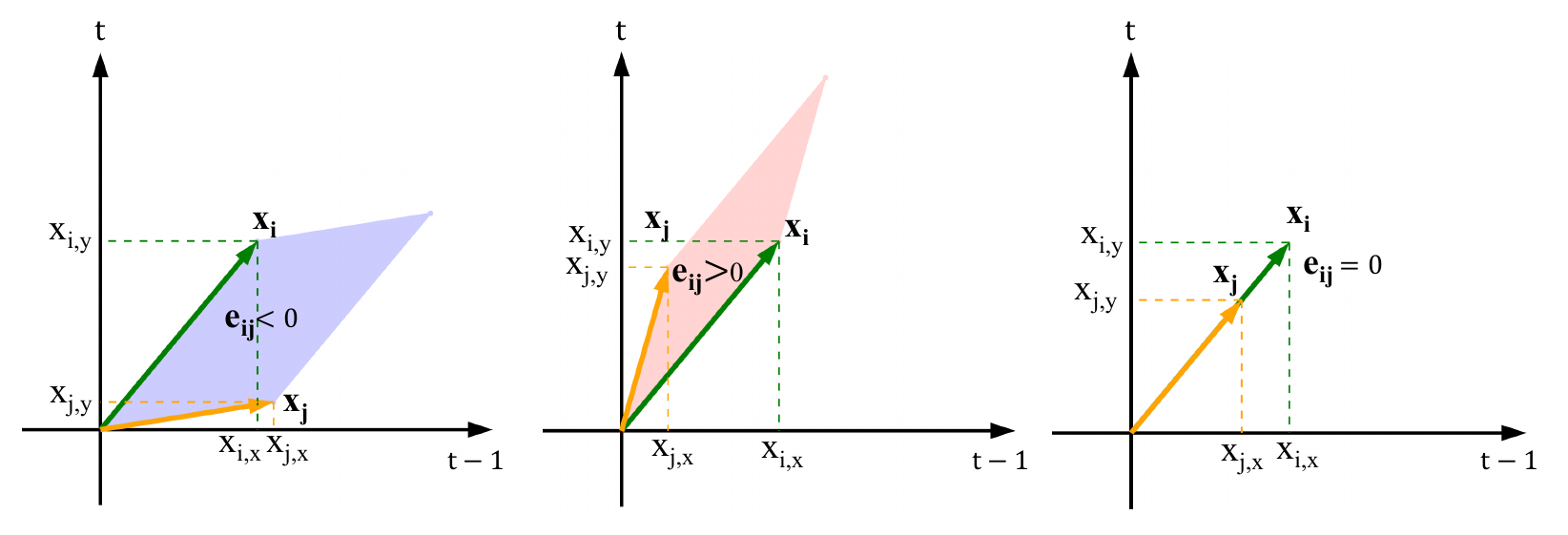} 
  \caption{\textbf{Geometric interpretation of the proposed edge signal.} Schematic representation of different configurations of vectors yielding positive (left), negative (center) and zero (right) edge signals. The edge represents the signed area spanned by the vectors $\vc{x}_i=(x_i [t-1], x_i[t])$ and $\vc{x}_j=(x_j[t-1], x_j[t])$.} 
  \label{fig:Edge_Signals}
\end{figure*}

We observe that Eq.~\eqref{eq:edge_signal_main} can be rewritten as the determinant of a $ 2 \times 2  $ matrix where the columns are two nodal signals $\vc{x}_i$ and $\vc{x}_j$, evaluated at time points $t$ and $t+1$:
\begin{equation}
\label{eq:edge_signal_determinant}
    \left|
        \begin{matrix}
        x_i[t-1] & x_j[t-1] \\
        x_i[t] & x_j[t]
        \end{matrix}
    \right|
\end{equation}
This representation admits a geometric interpretation as the signed area of the parallelogram spanned by the two temporal vectors $\vc{x}_{i,t} = (x_i[t-1], x_i[t])$ and $\vc{x}_{j,t} = ( x_j[t-1],  x_j[t])$, as presented in Fig.~\ref{fig:Edge_Signals}.

Intuitively, this determinant-based formulation highlights how the edge signal captures both the magnitude of the individual variations of the two nodes and their relative temporal alignment. In particular, it quantifies whether changes in one time series systematically align or misalign with those of the other over two consecutive time steps. The edge signal vanishes when the two vectors are linearly dependent, corresponding to perfectly aligned nodal dynamics. Moreover, its sign encodes the orientation of this temporal relationship, distinguishing which node tends to lead the other through the relative angle between them.

\begin{insightbox}{Connecting the Dots: Edge Signal and $2$-forms}{
    Fig.~\ref{fig:Edge_Signals} shows the signal representation in the $(t,t+1)$ space. The signed area admits a natural interpretation in terms of a \emph{wedge product}, an operator often used in differential goemetry which captures the oriented area spanned by the vectors $\vc x_{i,t}$ and $\vc x_{j,t}$. This is consistent with the determinant-based formulation presented in Eq.~\eqref{eq:edge_signal_determinant}, since the wedge product in $\mathbb{R}^M$ can be viewed as a coordinate-free generalization of the determinant in the framework of exterior algebra. Drawing from the language of differential geometry, the signal is closely related to a discrete 2-form in the sense of discrete exterior calculus, living on the $(t-1,t)$ space. We refer to \cite{crane2018discrete} for further details on wedge products and $k$-forms.
}
\end{insightbox}

\subsubsection{Link with Cyclicity Analysis}

Interestingly, the signal we discuss has strong links to quantities used in the context of cyclicity analysis \cite{baryshnikov2016cyclicity} and related applications \cite{abraham2024hemodynamic, maggs2025topology}. 

Cyclicity analysis, introduced in \cite{baryshnikov2016cyclicity}, provides a framework to infer lead–lag relationships in dynamical systems exhibiting cyclic interactions between time series at different nodes. Under additional assumptions on the underlying dynamics---such as sufficiently smooth trajectories and approximately cyclic temporal evolution---the interaction between nodes $i$ and $j$ can be characterized through a time-integrated measure of their joint evolution. This is obtained by constructing a lead–lag matrix $\vc M$, where the entry $(i,j)$ can be written as the time integral of a function of the two time trajectories $\vc x_i$ and $\vc x_j$ as follows \cite[Eq.~(3)]{abraham2024hemodynamic}:
\begin{equation}
    \label{eq:leadlag}
    [\ma M]_{i,j}= \frac{1}{2} \int_{t_1}^{t_2} x_i(t)\dot{x}_j(t)- x_j(t)\dot{x}_i(t) dt
\end{equation}
where $\dot{x}_i(t)$ indicates the derivative over time and can be, in practice, approximated with $x_i[t]-x_i[t-1]$, while the integral can be substituted by averaging. This highlights how the instantaneous formulation of the proposed signal in terms of discrete lags in Eq.~\eqref{eq:edge_signal_lag} corresponds to the above expression evaluated at a single lag, up to a scaling factor.

Such approaches have been used to characterize directional and cyclic structure in multivariate systems, including gene expression dynamics \cite{maggs2025topology} and brain activity patterns \cite{abraham2024hemodynamic,zimmerman2018dissociating}, to identify time propagation patterns.

Although some additional assumptions are required for cyclicity analysis (see \cite{baryshnikov2016cyclicity} or \cite{maggs2023simplicial,abraham2024hemodynamic}) that is beyond the scope of this work, these links provide further options to look at the proposed edge-level signal for capturing oriented dependencies in temporal systems. We believe that this connection with existing approaches further supports its relevance and opens the way for future theoretical and empirical investigations.

\subsubsection{Interpretation as Difference of Lagged Cross-Correlations}
\label{subsubsec:stationarity}
Although the edge signal in Eq.~\eqref{eq:edge_signal_main} is formulated in terms of instantaneous interactions, it can also be interpreted within the framework of stochastic processes, under additional assumptions. Specifically, we model the nodal time series as realizations of stochastic processes $\{X_i(t), t \in \mathbb{R}^{+}\}$ and $\{X_j(t), t \in \mathbb{R}^{+}\}$, where $X_i(t)$ and $X_j(t)$ are random variables taking values in $\mathbb{R}^{+}$. We impose that the two processes are Joint Wide-Sense Stationary (JWSS), which requires that the following conditions are satisfied simultaneously:
\begin{itemize}
    \item \textbf{$X_i$ and $X_j$ are wide-sense stationary.}\\
        For each process $\{X_m(t), t \in \mathbb{R}^{+}\}, \ m \in \{i,j\}$ it holds that $\mathbb{E}[X_m(t)] = \mu_m, \forall t, $ (mean functions do not depend on time), $\mathbb{E}[X_m^2(t)] < \infty$ (variances are finite), and   $\mathbb{E}[X_m(t) X_m(t+\tau)] = R_{m,m}(\tau)$ (autocorrelations depend only on lag). Furthermore, we assume $\mu_i, \mu_j = 0$ and $\sigma_i, \sigma_j = 1$.
    \item \textbf{Cross-correlations depend only on lag.}\\$
            \mathbb{E}[X_i(t) X_j(t+\tau)] = R_{i,j}(\tau), 
        $ which guarantees that the cross-correlations are independent of time. Note that, for the well-posedness of this condition, we also implicitly assume that cross-correlations are finite $\mathbb{E}[X_i(t)X_j(t+\tau)] < \infty$.
\end{itemize}

Under these assumptions, the edge signal can be studied as the realization of a stochastic process $\{E_{i,j}(t), t \in \mathbb{R}^{+}\}$. Its expected value can be interpreted as the difference between lag $1$ and $-1$ cross-correlations: 
\begin{equation}
    \label{eq:edge_signal_lagged_corr}
        \mathbb{E}[E_{i,j}(t)] =  R_{i,j}(1) - R_{i,j}(-1) 
\end{equation}
\begin{proof} Using linearity and JWSS, we find that
    \begin{equation*}
        \begin{aligned}
            \mathbb{E}[E_{i,j}(t)] 
            &=\mathbb{E}[X_i(t-1) X_j(t) -X_j(t-1) X_i(t)] \\
            &=\mathbb{E}[X_i(t-1)X_j(t)] - \mathbb{E}[X_j(t-1)X_i(t)]=\\
            &=\mathbb{E}[X_i(t-1)X_j(t)] - \mathbb{E}[X_i(t)X_j(t-1)]=\\
            &= R_{i,j}(1) - R_{i,j}(-1).
        \end{aligned}
    \end{equation*}
\end{proof}
The above formulation shows that, under mild regularity assumptions, the current signal can be interpreted as a stochastic measure of asymmetry between two nodal processes, considering a lag of $1$. For instance, assuming $R_{i,j}(1)$ and $R_{i,j}(-1)$ are both positive, the difference $R_{i,j}(1) - R_{i,j}(-1)$ quantifies a direction of temporal dominance, indicating whether fluctuations in $X_i$ tend to systematically precede (difference $>0$) or follow (difference $<0$) those in $X_j$ by one time point. 

This perspective further strengthens the interpretation of the proposed edge signal as an indicator of directional temporal dynamics, capturing time-dependent relationships between nodes communicating in the network.

\paragraph{Special Cases}
The interpretation of the edge process can be further clarified by considering a few illustrative scenarios. 
First, suppose that $X_i$ and $X_j$ are independent. In this case, one has $\mathbb{E}[X_i(t-1)X_j(t)] = \mathbb{E}[X_j(t-1)X_i(t)]=\mu_i\mu_j$, which implies that $\mathbb{E}[E_{i,j}(t)] = 0$. This reflects the intuitive idea that, in the absence of any statistical dependence, there is no expected systematic temporal asymmetry between the two processes. 
A second relevant situation arises when the lagged cross-correlation is symmetric: even in this case, $\mathbb{E}[E_{i,j}(t)] = 0$ as the two cross-correlations have the same value, indicating that fluctuations in $X_i$ and $X_j$ are equally likely to precede or follow each other at lag $1$.

\paragraph{Ergodicity and Implications for Computation} 
Assuming that for every $\tau$, the joint process $Y_{\tau}(t)=X_i(t)X_j(t + \tau)$ is mean-ergodic, temporal averages along a single observed realization can be used to estimate the expected value of the process. Indeed, under ergodicity, the time average of this realization converges almost surely (hence in probability) to the ensemble expectation:
\begin{equation}
    \bar{e}_{i,j} = \frac{1}{T}\sum_{t=0}\transpose e_{i,j}[t] \xrightarrow[t\to\infty]{\text{a.s.}} \mathbb{E}[E_{i,j}[t]] = R_{i,j}(1) - R_{i,j}(-1).
\end{equation}

Ergodicity guarantees that averaging over time along a single realization of the process is equivalent to averaging over an ensemble of independent realizations of the same process, which is particularly useful in practical applications where only one time series per node is available.

\subsubsection{Extension to Higher Dimensions}
Conceptually, the transition from a zeroth-order nodal signal to the proposed first-order edge signal can be understood as augmenting the nodal signal by leveraging the additional time dimension, in particular by considering two consecutive time points, $t$ and $t-1$. This construction naturally suggests a recipe for generalizing to higher-dimensional settings, where multiple consecutive time points are jointly considered.
For instance, in the three-dimensional case, a second-order signal (i.e., supported on triangles) can be defined as the determinant of a $3 \times 3$ matrix formed by node vectors $\vc{x}_i, \vc{x}_j, \vc{x}_k$ evaluated at three consecutive time instants $t, t-1, t-2$. Geometrically, this determinant measures the oriented volume of the parallelepiped spanned by the three vectors. From a statistical perspective, this construction can be interpreted as an antisymmetric combination of triple-wise lagged cross-correlations. While a detailed analysis of this higher-order signal lies beyond the scope of the present work, it provides a natural extension of the proposed framework to higher dimensions.

%% file: Main_text/Sections/3_Application.tex
\section{Case Study: \gls{tsp} to Study Brain Communication}
\label{sec:Application} 

\noindent In this section, we apply \gls{tsp} to brain imaging data. In particular, we showcase how \gls{tsp} approaches can be used to perform structure-informed analysis of brain communication.

We select this case study as a representative example in which the modelling challenges outlined in Section~\ref{sec:Domain_and_Edge_Signals} naturally arise, namely the definition of higher-order signals and domains.
In particular, brain processes involve interregional communication that can be conceptualized as time-varying information flow between two or more regions, reflecting their coordinated activity. However, state-of-the-art neuroimaging techniques typically provide information about static pairwise connections between regions, often represented as a graph, together with indirect measurements of brain activity, modeled as nodal signals \cite{preti2023graph,huang2018graph}. Consequently, higher-order structures and signals are not directly observed, but inferring them from these lower-order representations may help provide a more informative characterization of brain organization and function.

From an applied perspective, we are also motivated by the increasing attention to the role of high-order interactions \cite{santoro2023higher, varley2023multivariate, luppi2023information} and topology-driven descriptors \cite{santoro2024higher, varley2025topology, expert2022higher} of brain structure and communication mechanisms; see also \cite[Sect. 6]{abiad2026hypergraphs}. Interestingly, a few recent works have explored the use of \gls{tsp} for neuroimaging analysis \cite{bispo2026multimodal, sardellitti2025topological, Santoro2025EdgeLaplacians, roy2025hodge}. While some of them have moved beyond the preliminary stage and begun to offer more unified perspectives and \gls{tsp} pipelines \cite{bispo2026multimodal}, a shared, well-established framework for \gls{tsp}-driven neuroimaging analysis is still lacking. In this context, our case study illustrates how the edge signal introduced in Sec.~\ref{sec:Domain_and_Edge_Signals}, capturing lead-lag dynamics between couples of regions, can be used to explore brain communication processes, highlighting the potential of \gls{tsp}-informed analyses for applications.

We emphasize that, while the context and interpretation of this case study are specific to neuroimaging, the underlying rationale and analysis are broadly applicable to a wide range of domains. The role of the case study is therefore not merely illustrative, but methodological: it provides practitioners with a concrete example of how these tools can be implemented and interpreted in practice on real-world irregular domains,which remains one of the most delicate aspects in the application of TSP methods. For this reason, we deliberately frame the analysis in a way that highlights its broader applicability whenever possible. For instance, Fig.~\ref{fig:brain_eigenvectors} illustrates, through a concrete example, how spectral modes of the Hodge Laplacian on irregular domains can be interpreted. To the best of the authors’ knowledge, this is the first visualization of Hodge-Laplacian eigenvectors in a neuroimaging setting that accounts for the orientation of the underlying brain simplicial complex.

\subsection{Neuroimaging Background}
\label{subsec:neuroimaging_preliminaries}
A longstanding question in neuroscience concerns how brain organization and anatomical structure shape communication across neural population, ultimately supporting complex functions ranging from sensation, motor control and cognition \cite{fotiadis2024structure, suarez2020linking, preti2019decoupling, liegeois2020revisiting}. Neuroimaging, and in particular \gls{mri}, allows this question to be investigated non-invasively in humans at the level of brain regions, providing complementary insights about brain structure and function.

On the structural side, \gls{dwi} is an \gls{mri}-based technique that measures how water molecules diffuse along different directions in brain tissue \cite{bammer2003basic}. This orientation sensitivity provides information about the density and orientation of white matter fibers connecting pairs of brain regions. These fibers form the anatomical pathways that support long-range communication across the brain and are modeled with the so-called \gls{sc} matrix \cite{zhang2022quantitative}. Considering $N$ brain regions, the \gls{sc} is typically expressed as a matrix $\mathbf{A} \in \mathbb{R}^{N \times N}$ \cite{behrens2012human}, where each entry represents the white matter connection between a pair of regions. Entries can be weighted according to fiber density or thickness, or thresholded to retain only the strongest connections, yielding a sparse binary matrix.

On the functional side, \gls{fmri} provides an indirect measure of regional neuronal activity by tracking changes in the \gls{bold} signal over time \cite{ogawa1990brain}. For a single subject, one obtains nodal signals $\vc{x}^{(0)} \in \mathbb{R}^{N \times L}$, where $L$ indicates the number of timepoints.  Regions can also be grouped into larger-scale \emph{brain networks}, which consist of sets of regions that tend to work together and support similar functions, providing a more parsimonious and functionally meaningful representation of brain organization \cite{yeo2011organization}. 

\subsection{Research Question}
\label{subsec:research_question}

Brain regions that are strongly connected via white-matter pathways often show correlated functional activity, a relationship known as \emph{structure–function coupling} \cite{fotiadis2024structure, preti2019decoupling, park2013structural}. This coupling reflects how anatomical connectivity shapes interregional brain dynamics, and has been investigated across different conditions and tasks (see, for instance, \cite{liu2022time}).

Notably, \gls{gsp} provides a framework for jointly analyzing brain structure and function by representing the \gls{sc} as a brain graph and \gls{fmri} regional activation patterns as nodal signals \cite{huang2018graph}. Standard \gls{gsp} approaches, however, have limitations in capturing the direction of communication between regions: they rely on static, undirected graphs and primarily reflect diffusive signal dynamics with no directionality, rather than explicit information flows \cite{huang2018graph, seguin2023brain}. Directed extensions exist, but they treat directionality as a fixed structural limiting their ability to capture the time-varying, bidirectional interactions observed in the brain.

Motivated by these limitations, we investigate whether node-derived edge signals can provide structure-informed insights on brain communication. The goal of this analysis is twofold. First, we aim to determine whether pairs of brain regions exhibit consistent patterns of lead-lag dynamics across subjects. Second, we aim to to evaluate whether, and to what extent, these patterns can be attributed to irrotational, divergence-like or circulant, curl-like brain activity patterns, and which brain networks are involved.

\begin{figure*}[!b] 
    \centering
    \includegraphics[width=\linewidth]{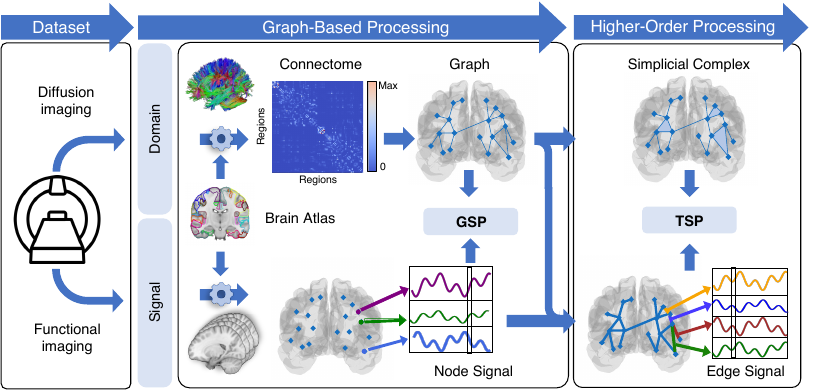} 
    \caption{
        \textbf{Schematic of the dataset structure and processing.} From left to right: magnetic resonance imaging (MRI) diffusion and functional imaging data are first obtained for every subject. On the domain side, diffusion imaging data are then preprocessed to extract the tractogram, which includes all the white matter fibers connecting different brain voxels. Then, a brain atlas is used to average the fibers connecting each pair of brain regions, resulting in an $N \times N$ matrix (structural connectome). The connectome is averaged across subjects, thresholded at the $90$th percentile, and binarized to construct a brain graph. A simplicial complex is then built as the $3$-clique complex of the brain graph. On the signal side, functional imaging data are preprocessed and segmented into regions using the same brain atlas, obtaining a nodal signal over the $N$ regions that is bandpass filtered in the range 0.1–1 Hz and z-scored over time. Then, the edge signal is computed on the edges of the previously constructed brain graph, as defined in Eq. \eqref{eq:edge_signal_main}. Finally, the simplicial complex and the edge signal are jointly processed leveraging \gls{tsp}.
    }
    \label{fig:pipeline}
\end{figure*}

\subsection{Methods}

\subsubsection{Dataset and Preprocessing}
The dataset used is freely available from the \gls{hcp} 
\cite{van2013wu, van2012human}, from which we obtain structural and functional neuroimaging data from $100$ unrelated participants ($54$ females and $46$ males, mean age = $29.1 \pm 3.7$ years). The processing pipeline is illustrated in Fig.~\ref{fig:pipeline}.

We consider the \gls{fmri} data at resting state (i.e., participants were not engaged in any explicit task), with a \gls{tr} of $0.72$ s. 
The data had already undergone minimal preprocessing prior to download \cite{schaefer2018local}; additional preprocessing was then performed following standard steps described in \cite{van2021makes}\footnote{In particular, involving volume realignment, regression of nuisance signals (scanner-related linear and quadratic drifts, motion regressors and their first derivatives, white matter, cerebrospinal fluid signals, and their first derivatives), and spatial smoothing (FWHM=5mm)).}. The time series were then bandpass-filtered in the $0.01-0.10$ Hz range to remove non-neural components, such as cardiac or respiratory artifacts. Lastly, voxel-wise fMRI time series were averaged within reference atlas brain regions and z-scored. Since two recordings of $1200$ \gls{tr} were available for each subject, the resulting normalized time series were subsequently concatenated into a single individual run. A standard atlas of $100$ regions was considered \cite{schaefer2018local}, resulting in a network of $100$ nodes. For analyses at a broader scale, these regions were further grouped into seven well-known large-scale brain networks (the canonical Yeo networks \cite{yeo2011organization}).

\subsubsection{Deriving the Simplicial Complex from the \gls{sc} Graph}
First, we derive the brain matter graph from white-matter connectivity following a threshold-based approach. The structural connectome is defined such that each edge between pairs of regions corresponds to the mean number of white matter fibers for each subject. Connectivity matrices are then averaged across subjects. This step ensures that the same domain is used for all subjects, thereby enabling the analysis of signals on a common underlying topology. The resulting mean connectome is thresholded by retaining only the top 10\% strongest connections and subsequently binarized, resulting into an unweighted graph. Then, a simplicial complex of dimension $2$ is built as the $3$-clique complex of the brain graph. 

This approach results into a simplicial complex with $N=100$ nodes (brain regions), $E=495$ edges, and $T=720$ triangles, with boundary matrices having dimension $\mathbf{B}_{1} \in \mathbb{R}^{100 \times 495}$ and $\mathbf{B}_2 \in \mathbb{R}^{495 \times 720}$.
The analysis of the Hodge Laplacian $\mathbf{L}_1$ reveals that $\operatorname{dim}(\operatorname{im}(\mathbf{L}_{1\downarrow}))=99$, $\operatorname{dim}(\operatorname{im}(\mathbf{L}_{1\uparrow}))=392$, and $\operatorname{dim}(\operatorname{ker}(\mathbf{L}_1))=4$, indicating the presence of 4 empty cycles represented by unfilled groups of $3$ or more edges.

\subsubsection{Deriving the Edge Signal from fMRI Nodal Time Series}
Starting from the original \gls{fmri} time series, the edge signal is computed over all pairs of connected nodes in the simplicial complex as in Eq.~\eqref{eq:edge_signal_main}, adopting the ordering convention $e_{i,j}=[v_i,v_j], \ i<j$. Note that, due to the lag, the edge time series for each pair of regions has length of one time point less than the original nodal ones, with a length of $L=2399$ \gls{tr} ($2399 \cdot 0.72 = 1727.28$ s), and a final dimension of $\vc{X}^{(1)}_{k} \in \mathbb{R}^{E \times L}$ for each subject $k$. $\bar{\mathbf{X}}^{(1)} \in \mathbb{R}^{E}$ corresponds to is its average over time.

The edge components ${\mathbf{X}}^{(1)}_{k\downarrow}, \mathbf{X}^{(1)}_{k\uparrow} \in \mathbb{R}^{E \times L}$ are matrices of all edge timeseries for subject $k$ resulting from the projection of $\mathbf{X}^{(1)}_{k}$ onto the space of $\llow$ and $\lup$, respectively, as illustrated in Fig.~\ref{fig:ExampleIncidenceMatrices} d), f).
Finally, the divergence and the curl over the average edge signal are computed for each subject $k$ as
$\operatorname{div}(\bar{\ma{X}}^{(1)}_k)=\mathbf{B}_1 \bar{\ma{X}}^{(1)}_k \in \mathbb{R}^{N}$
and $\operatorname{curl}(\bar{\ma{X}}^{(1)}_k)=\mathbf{B}_2\transpose \bar{\ma{X}}^{(1)}_k \in \mathbb{R}^{T}$,
respectively.

\subsection{Results}

\begin{figure*}[htbp]
    \centering \includegraphics[width=\linewidth]{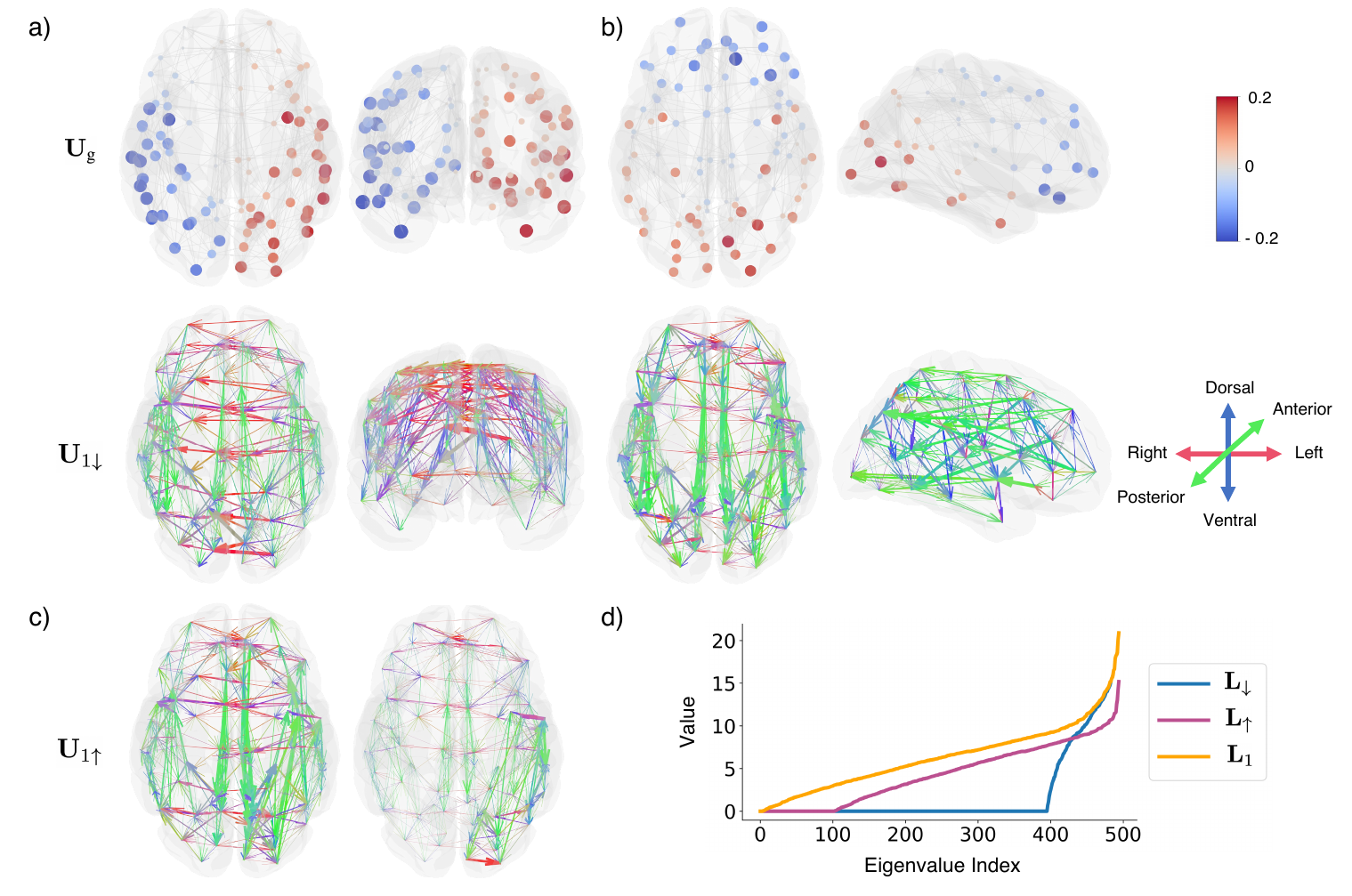} 
    \caption{\textbf{From GSP to TSP: Different perspectives on brain connectivity eigenvectors.} a) Different views of the first nonzero eigenvector from the graph Laplacian $\mathbf{L}$ and lower component of the Hodge Laplacian (edge Laplacian) $\mathbf{L}_{1\downarrow}$, which are associated with the lowest frequency of variation over the domain. Arrow thickness is proportional to the magnitude of the eigenvector on the corresponding edge. Note that each arrow reflects the actual flow direction along each edge, as the sign of the corresponding eigenvector entry has been adjusted with respect to the chosen reference orientation. The graph eigenvector captures smoothly varying patterns and is associated with a gradient between the two brain hemispheres. The edge-based perspective highlights how this variation captures a latero-lateral flow from one hemisphere to the other, as indicated by the red arrows, which exhibit a consistent direction. b) Second eigenvector of the graph and the edge Laplacian. The nodal perspective highlights a contrast between anterior and posterior regions of the brain, whereas the edge eigenvector captures a coordinated gradient of edge flow from anterior to posterior regions of the brain. c) Two eigenvectors of $\mathbf{L}_{1\uparrow}$ associated with different frequencies: the eigenvector in the figure on the left exhibits a lower-frequency, more spatially distributed variation pattern, whereas the figure on the right shows an eigenvector with higher-frequency, exhibiting a more localized pattern. Note that there is no association between the eigenvectors of $\mathbf{L}_{1\uparrow}$ and the ones of the graph Laplacian, as the former depend on the triadic (circulant) interactions that lie in the kernel of the graph Laplacian. d) Eigenvalues of the Hodge Laplacian $\mathbf{L}_1$ and its components $\mathbf{L}_{1\downarrow}$ and $\mathbf{L}_{1\uparrow}$.}
    \label{fig:brain_eigenvectors}
\end{figure*}

\subsubsection{Spectral Insights on the Simplicial Complex}
We first carry out a spectral decomposition of the domain to illustrate structural modes of variation of the simplicial complex. Specifically, we perform the spectral analysis of the Hodge Laplacian as discussed in Sec.~\ref{subsec:Fourier}, and compare some elements of this Fourier basis with the ones obtained from the graph Laplacian $\mathbf{L}$, shown in  Fig.~\ref{fig:brain_eigenvectors}. A first observation from the figure is that edge-based representations change the semantics of the spectral domain from node-centric to connection-centric activity patterns. In this setting, spectral modes describe structured variations over the edge domain, capturing interaction dynamics between regions rather than localized activity. This perspective may be particularly relevant for studying communication processes such as brain ones, as it directly represents quantities encoding some form of information flow between brain regions. This visualization also provides a solid basis for improved interpretability in the spectral analysis of higher-order oriented signals. Indeed, although beyond the scope of this work, projecting signals onto individual eigenvectors, for instance to create an embedding in a lower space or to filter specific frequencies, acquires here a visual interpretation: large positive (resp. negative) coefficients indicate alignment with (resp. opposition to) the corresponding spectral mode, while their magnitude reflects the amount of signal energy captured by that component. Importantly, while the underlying concept is analogous to classical GSP, one key difference is that orientation is intrinsically encoded in the spectral modes.

Focusing on the different types of eigenvectors, Fig.~\ref{fig:brain_eigenvectors} a), b) illustrates how those of $\mathbf{L}_{1\downarrow}$ relate to their graph counterparts, as also discussed in Sec.~\ref{subsec:gsp_tsp}, giving rise to gradient-like edge patterns. Interestingly, looking at the first 2 eigenvectors of $\ma L_{1\downarrow}$ reveals that they exhibit organized variations along the two principal brain axes (Fig.~\ref{fig:brain_eigenvectors}a), b)). On the other hand, $\mathbf{L}_{1\uparrow}$ encodes information associated with cyclic patterns of flow that have no direct analogue at the level of the graph Laplacian, as they correspond to components of the signal lying in the kernel of the lower Laplacian. 

\subsubsection{Edge Signal Contributions From Up/Down Subspaces}
Before proceeding with the analysis, an important preliminary step is to assess whether circulations are present in the signal; i.e., whether a non-negligible portion of the signal lies in the image of $\mathbf{L}_{1\uparrow}$. To this end, similarly to previous works  \cite{barbarossa2020topological, bispo2026multimodal}, we project all edge signals onto the subspaces associated with $ \ma L_{1\downarrow}$ and $ \ma L_{1\uparrow}$ and compute the corresponding energies. Since the subspaces deriving from the Hodge decomposition are orthogonal, the sum of the energies of the lower and upper components equals the total energy of the edge signal, excluding the contribution from the harmonic component. For each subject $k$, we define the proportion of curl energy $p_k$ as
\begin{equation}
    p_k = \frac{\|\mathbf{X}^{(1)}_{k \uparrow}\|_F^2}{\||\mathbf{X}^{(1)}_{k \downarrow}|\|_F^2+ \|\mathbf{X}^{(1)}_{k \uparrow }\|_F^2},
\end{equation}
and use this value to assess the amount of energy in the higher-order space. On average, $p_k= 0.25 \pm 0.074$ (mean ± standard deviation computed across subjects), thus about a quarter of the energy of the edge signal corresponds to circulations around $2$-simplices.
This justifies continuing the analysis on the simplicial complex of dimension $2$, incorporating higher-order components of the domain.

\begin{figure*}[htbp]  
    \centering
    \includegraphics[width=\linewidth]{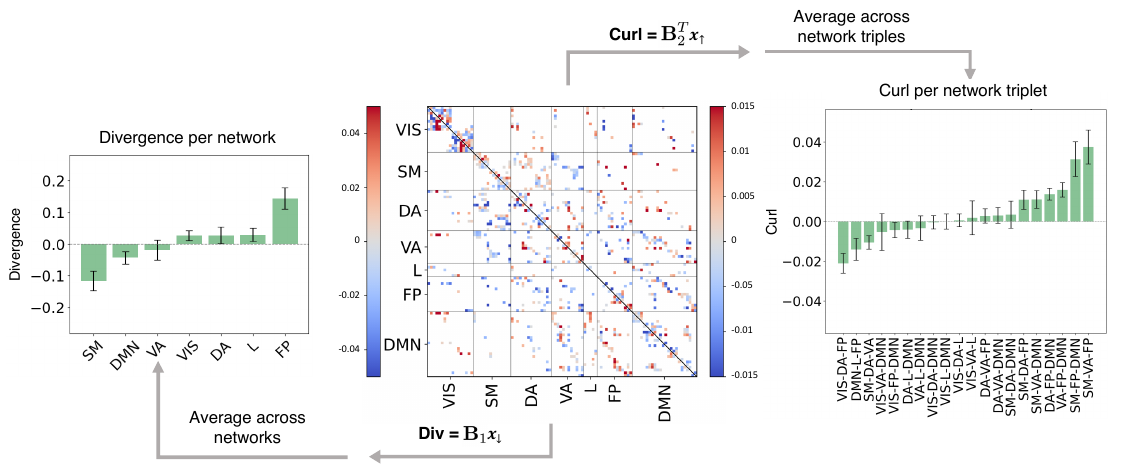}
    \caption{\textbf{Brain edge signals and projections on the subspace of the Hodge Laplacian.}
    Center: matrix showing the mean edge signal (across time and subjects) for each pair of brain regions, with ordering and grouping based on brain networks. The lower triangular matrix shows $\bar{\ma X}^{(1)}_{\downarrow}$, i.e., the projection of the average edge signal onto the space of $\mathbf{L}_{\downarrow}$, whereas the upper triangular matrix shows $\bar{ \ma X}^{(1)}_{\uparrow}$, i.e., the projection onto the space of $\mathbf{L}_{\uparrow}$. To ease interpretation, edges are reorganized and shown symmetrically, with different color scales to compensate for magnitude difference. In particular, for each entry $i,j$ in the lower triangular matrix, a positive signal $\bar{x}^{(1)}_{i,j}$ indicates a flow from region $i$ to region $j$; in the upper triangular one, the same direction of flow is indicated by a positive value of the corresponding entry $\bar{x}^{(1)}_{j,i}$.
    Left: bar plots representing the the average divergence over time, aggregated by brain networks. Mean $\pm 3$ standard errors across subjects are reported. Positive divergence indicates that the related network mostly receives flow and acts as a sink (i.e., mostly follows), whereas negative divergence indicates the opposite.
    Right: bar plots representing the temporal average of the curl, aggregated by triplets of networks. Positive curl values indicate consistency with the reference ordering induced by the network labels (e.g., a positive value of SM-VA-FP indicates a circulation SM $\rightarrow$ VA $\rightarrow$ FP, up to even permutations), whereas negative values correspond to to circulations that are opposite to the reference ordering. Brain network legend: VIS: Visual, SM: Sensorimotor, DA: Dorsal Attention, VA: Ventral Attention, L: Limbic, FP: Frontoparietal, DMN: Default Mode.}
  \label{fig:Brain_Edge_signal}
\end{figure*}

\subsubsection{Edge Signal Analysis}
We now turn to the analysis of the different edge components based on the lead-lag dynamics between pairs of regions, and their interpretation. As we work with resting-state data, we can assume that the edge signal satisfies the assumptions introduced in Sec.~\ref{subsubsec:stationarity}. In this context, the temporal mean of each edge signal for a given subject provides an estimate of the difference between lag-one cross-correlations. Therefore, we compute the temporal average of each component of the edge signal for each subject $k$, obtaining $\bar{x}^{(1)}_{i,j,k}$ as described in Eq.~\eqref{eq:edge_signal_main}. Before discussing its interpretation, we analyze the two components of the signal in terms of lag-$1$ cross-correlations.

As discussed in the previous section, we consider the estimated lagged cross-correlations for a single subject (for ease of notation, we omit the index $k$): $\hat{r}_{i,j}(1) = \frac{1}{L-1} \sum_{t=1}^{L-1} x_i^{(1)}[t-1] x_j^{(1)}[t], \quad \hat{r}_{i,j}(-1) = \frac{1}{L-1} \sum_{t=1}^{L-1} x_j^{(1)}[t-1] x_i^{(1)}[t].$ When both quantities are positive, the average edge signal
$\bar{x}_{i,j}^{(1)} = \hat{r}_{i,j}(1) - \hat{r}_{i,j}(-1)$ captures the direction of temporal precedence between regions: positive values indicate that region $i$ tends to lead region $j$ at one time lag, while negative values indicate the opposite. We find that $\hat{r}_{i,j}(1)$ is positive in $98.92 \pm 1.67\%$ of cases, while $\hat{r}_{i,j}(-1) $is positive in $98.90 \pm 1.72\%$ of cases (mean $\pm$ standard deviation across subjects), supporting this interpretation.

For each subject $k$ and each existing edge $e_{i,j}$ we compute the time-averaged lower signal component $\bar{x}^{(1)}_{i,j,k \downarrow} \in \mathbb{R}^{E}$, which can be attributed to gradients of flows between nodes, and the respective upper edge components $\bar{x}^{(1)}_{i,j,k \uparrow}\in \mathbb{R}^{E}$, related to local internodal circulations.

We then further average these signals across subjects to capture group-level trends for each edge. The results are shown in the central plot of Fig.~\ref{fig:Brain_Edge_signal}. We observe subtle but observable differences between the upper and lower components across region pairs, indicating that certain dynamical features captured by the upper component are not explained by the lower one, and vice versa. Some region pairs, and their corresponding brain networks, exhibit strong contributions from both components. For instance, edges between regions of the Visual Network (VIS), which is responsible for the relay and processing of visual information, show higher absolute values in both signal components, as illustrated in the corresponding entries of the central plot of Fig.~\ref{fig:Brain_Edge_signal}.

\subsubsection{Curl and Divergence Analysis}
Because net signal in- or outflow is naturally captured at the level of nodes, whereas processes involving circulations around triplets of nodes are better described at the triadic scale, we compute divergence on nodes and curl on the triangles of the brain simplicial complex as described in the Methods. These quantities are then aggregated to obtain measures at the brain network level. Specifically, we compute the average divergence for individual networks and the average curl for triplets of networks. Note that, although the proposed edge signal can only be interpreted in terms of lagged time relations and does not imply causality, we refer to a region which tends to anticipate another as “sending,” and the one which tends to follow as “receiving” reflecting their role in the signal flow.

When examining divergence over individual networks, the Sensorimotor Network (SMN), responsible for planning, control, and execution of voluntary movements, as well as the processing of somatosensory information from the body, emerges as showing the strongest negative divergence, indicating a predominantly sending role (Fig.~\ref{fig:Brain_Edge_signal}, left). In contrast, some networks tend to show the opposite behavior, exhibiting positive divergence which reflects a more receiving-oriented activity. This is particularly evident in the Frontoparietal Network (FPN), a set of brain regions involved in high-level functions such as attention, decision-making, and problem solving. Interestingly, this sending-receiving (or, equivalently, anticipating-following) gradient is largely consistent with previous work studying the directionality of brain communication through asymmetry measures computed on the structural connectivity on similar data \cite{seguin2019inferring}.

When examining general trends in the curl (Fig.~\ref{fig:Brain_Edge_signal}, right), several network triplets exhibit coordinated triple-wise patterns, with particularly strong involvement of the FPN and the Default Mode Network (DMN), associated with self-referential processes and memory and often active in the absence of tasks. Investigating interactions from this perspective is a new and exciting possibility that can further inspire network neuroscience. 

\subsubsection{Sign-Flip Test on the Divergence}
\begin{figure*}[htbp]
    \centering
    \includegraphics[width=\linewidth]{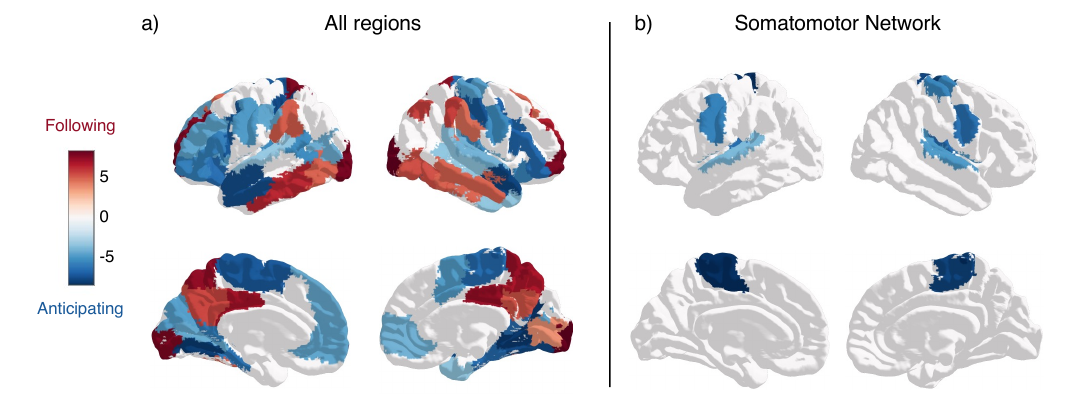}
    \caption{\textbf{Divergence over brain regions}. a) Regions surviving the nonparametric sign-flip test on the mean (p-value $\leq$ 0.05, Bonferroni-corrected), colored by the empirical z-score of the mean with respect to the null distribution $\hat{z}= \frac{\mu 
    - \mu_\text{perm}}{\sigma_\text{perm}}$. Positive average divergence (red) and negative average divergence (blue) indicate that the region is consistently following or anticipating the ones it is connected to by a lag of $1$ time point. b) Regions surviving the test belonging to the Sensorimotor Network, which exhibit negative divergence.}
  \label{fig:Divergence}
\end{figure*}

We then analyzed the divergence at the level of individual brain regions, to disentangle the possible presence of more nuanced patterns otherwise averaged out at the network level. To identify regions that consistently show positive or negative divergence across subjects, we performed a nonparametric sign-flip test on the mean ($n_{\text{permutations}} = 20{,}000$). Fig.~\ref{fig:Divergence} a) shows the empirical z-scores of the brain regions that survived the test ($p < 0.05$ after Bonferroni correction), revealing largely symmetric patterns between hemispheres, consistent with known principles of brain organization and physiology. 

Looking in particular at the SMN  (Fig.~\ref{fig:Divergence} b)), we find that most regions in both hemispheres show a negative divergence indicating a sender (i.e., anticipating) role, in agreement with their average shown in the bar plot of Fig.~$\ref{fig:Edge_Signals}$ and with previous work \cite{seguin2019inferring}.

%% file: Main_text/Sections/4_Discussion.tex
\section{Discussion}
\label{sec:Discussion} 

\noindent We summarize here fundamental insights on \gls{tsp}, highlighting promising research directions and key limitations. We then briefly discuss the computational challenges of the framework.

\subsection{Concluding Remarks on the Essence of TSP}

\subsubsection{The Role of Network Topology and Geometry}
\gls{tsp} methods based on the Hodge Laplacian and related operators enable signal analysis by simultaneously accounting for topology and geometry of the underlying network, encoded in the Hodge Laplacian and its spectral representation. 

On the one hand, simplicial complexes are constructions arising from applied topology and are therefore naturally equipped to capture topological invariants of the underlying space, such as high-dimensional holes, which lie in the kernel of the Hodge Laplacian. In \gls{tsp}, these invariants are reflected in the harmonic components of the signal induced by the Hodge decomposition. 

On the other hand, the non-harmonic components encode the local relational structure of the network, captured by the image of the Hodge Laplacian. In particular, the upper and lower components of the operator encode neighborhood relationships across multiple scales via combinatorial analogues of classical vector calculus operators, such as curl and divergence, and notions from differential geometry.  Importantly, while these local relationships can be viewed as defining the ``geometry'' of the network, we emphasize that they arise purely from combinatorial information, depending only on network connectivity and not requiring an embedding in a physical or metric space, although such embeddings may also be introduced.

This dual nature of the Hodge Laplacian enables a principled representation of both global structural properties and local relations across multiple scales and for relational data of any arbitrary order, which is one of the key strengths of the \gls{tsp} framework.

\subsubsection{Higher-Order Domains and Signals} 
A recurring theme in \gls{tsp} literature is the ability to handle “higher-order” domains and signals. However, the interpretation of this notion varies across contexts, potentially leading to ambiguity that we addressed in the previous sections.  

In general, higher-order domains and representations are suited to capture multi-way interactions or dependencies in data. In the context of this work, these domains take the form of simplicial complexes. In several cases, however, more general topological domains can be considered, depending on the dataset and the application at hand. For instance, cell complexes provide a more flexible framework for modeling higher-order interactions, as they allow polygonal-like structures while still enabling the use of Hodge-theoretic tools ~\cite{roddenberry2022signal, sardellitti2021topological, hoppe2025don}. For this reason, cell complexes have been used to model water distribution systems~\cite{cattai2025leak, cattai2025leak}, traffic networks~\cite{sardellitti2024topological}, and brain connectivity~\cite{bispo2026multimodal}. More broadly, the \gls{tsp} framework can be generalized further with the theory of sheaves, which provides a unifying language for signal analysis over general topological spaces~\cite{hatcher2001algebraictopology, battiloro2024tangent, robinson2014topological}. While this line of research lies beyond the scope of the current work, it highlights promising directions for future developments at the intersection of topology and signal processing. 

Furthermore, we have introduced higher-order signals as the ones living on higher-order simplices, including network edges as a first instance. In different sections, we have discussed how higher-order signals particularly suited to \gls{tsp} are those endowed with a notion of orientation and directionality. This is a subtle yet crucial aspect of the analysis, closely related to the interpretation of these signals as discrete analogues of $k$-forms in differential geometry. Moreover, while \gls{tsp} lends itself to the analysis of signals of all orders and is in principle applicable to nodal (i.e., zeroth-order) signals, it becomes particularly insightful and clearly differentiates from \gls{gsp} when extended to higher-order signals, as zeroth-order signals do not have an orientation.

\subsection{Considerations on Derived Domains and Signals}
In Sec.~\ref{sec:Domain_and_Edge_Signals}, we addressed the problem of inferring higher-order domains and signals from lower-order ones. This active area of research is fundamental for fostering cross-talk between communities and advancing \gls{tsp} and its applications across disciplines. Indeed, datasets are often not provided in a form directly suitable for \gls{tsp}, yet with appropriate preprocessing this framework could still be meaningfully leveraged to extract useful insights.

Beyond providing illustrative classifications of approaches for deriving higher-order domains and signals, we highlighted how, in signal processing contexts, the two problems are often deeply intertwined. For example, we discussed how introducing higher-order simplices makes it possible to capture circulations around them by projecting the signal onto the solenoidal space, whereas they would otherwise lie in the kernel of the Hodge Laplacian. Similarly, the introduction of $2$-simplices enables a more nuanced notion of signal smoothness that can be used to penalize certain circulation patterns, as discussed in detail in \cite{schaub2021signal}.

We then briefly examined a few examples of edge signals derived from nodal signals. Specifically, we showed how to derive a signal living on the edges of a simplicial complex that captures a network-aware representation of pairwise nodal lagged dynamics, providing both a geometric interpretation and an alternative perspective grounded in basic concepts of stochastic processes. In its deterministic formulation, the signal requires no additional assumptions nor external parameters and captures instantaneous dynamics, reflecting pairwise relationships between nodal temporal trajectories. Alternatively, by introducing suitable statistical assumptions, the same construction can be interpreted as a difference of lagged cross-correlations between random processes. Either perspective can be leveraged, depending on the objectives of the analysis and the assumptions imposed on the data. Furthermore, we have highlighted its connection with a similar approach leveraged in cyclicity analysis to infer lead-lag relationships in simple dynamical models. An important caveat, however, concerns the interpretation of the measure. Indeed, it only relates to notions of anticipation or delay and provides no evidence of causal influence from one node to another, as it may also arise from spurious dynamical patterns. Beyond this specific case, we hope that the underlying rationale behind the proposed construction can inspire the identification or derivation of meaningful signals in a broad range of different settings.

\subsection{Case Study on Brain Imaging Data}
In Sec.~\ref{sec:Application}, we presented a case study on \gls{tsp} for brain imaging data, driven by the increasing importance of higher-order interactions and edge-centric perspectives in network neuroscience \cite{betzel2023living}. First, we constructed a simplicial complex from brain white-matter data and computed edge signals from functional brain time series during resting state, using the scheme introduced before that captures lagged relationships. We then analyzed the different components of the edge space using the Hodge Laplacian and the curl and divergence operators. The results of this proof-of-concept illustrate how simple \gls{tsp}-based processing can uncover lead–lag relationships embedded in the signals and shaped by their underlying topological structure. Here, we briefly discuss choices related to domain modeling and the Hodge Laplacian---both of which may generalize to other datasets---and the interpretation of the main results.

While the lower-order components of the signal depend only on the edges of the graph, the higher-order ones are highly affected by the amount and localization of $2$-simplices. In this work, we chose to consider the $3$-clique complex of the brain graph and hence include all possible $2$-simplices; however, alternatively, more data-driven inference approaches could also have been employed, such as those mentioned in Sec.~\ref{par:domain_geometry}. 
Furthermore, while we relied on the unweighted combinatorial Hodge Laplacian, other options include the weighted version of it or  the random walk Hodge Laplacian, which normalizes weights by taking into account the degree distribution of the edges \cite{schaub2020random}. Comparing results obtained with these operators could be useful to relate contribution of different normalization schemes to the signal components. 

One first interesting result was that a non-negligible fraction of the signal energy lies in the rotational space, reflecting cyclic interaction patterns that could only be captured by introducing higher-order relationships in the structure of the simplicial complex, thereby justifying a \gls{tsp}-driven analysis. Furthermore, decomposing the edge-based signal into lower and upper components revealed that lead–lag relationships varied across region pairs and networks, with each pair usually exhibiting distinct behavior across different edge subspaces.

Aggregating the divergence at network-level highlighted a heterogeneous involvement of brain regions that consistently act as sources or sinks; i.e., regions that systematically anticipate or follow others. Most interestingly, those results are largely consistent with previous studies aimed at identifying mostly sending and receiving brain regions, based on structural connectivity measures alone~\cite{seguin2019inferring}. Furthermore, beyond divergence-driven effects, we observed that the curl-related component is slightly more pronounced in networks associated with higher-order cognitive processing. Future work could focus on specific signal components to examine their contributions across frequency bands and Hodge subspaces. Several open questions remain, including validation across datasets, reproducibility, and the relationship with established measures of brain function such as functional connectivity \cite{hutchison2013dynamic}.
Addressing these issues will be essential to assess the robustness of this proof-of-concept study.

In any case, we emphasize that \gls{tsp} provides a powerful alternative for investigating the dynamics of structure–function coupling in the brain---a longstanding quest in neuroscience---through the perspective of brain communication \cite{preti2019decoupling, seguin2023brain}. Compared to other methods, such as dynamical causal modeling (DCM)\cite{friston2014dcm} or Hidden Markov Models (HMMs) \cite{quinn2018task}, which have been widely used to study directed and latent brain dynamics, \gls{tsp} enables the analysis of time-varying, directed communication in a network-informed manner. Unlike traditional \gls{gsp} approaches, it naturally captures dynamic directional interactions, potentially providing new insights into how brain structure constrains and informs its communication.

\subsection{Considerations on Computational Complexity}
From an applied viewpoint, a significant limitation of \gls{tsp} lies in its computational complexity. Indeed, for a fixed number of vertices, the number of $k$-simplices grows combinatorially with $k$, leading to a rapid increase in the size of the simplicial complex as higher-order simplices are included.

Signal processing techniques may help mitigate the associated computational burden. In particular, several approaches have been proposed to obtain sparse signal representations that facilitate efficient processing, including generalizations of wavelets \cite{roddenberry2022hodgelets} and dictionary learning methods \cite{grimaldi2025topological} for signals defined on simplicial complexes. An even more topology-informed direction to address this challenge involves combining signal processing methods with discrete Morse theory, a branch of computational topology \cite{forman2002user}. In particular, in a recent work, Ebli \emph{et al.} \cite{ebli2024morse} have presented a way to reduce the dimension of signals living on simplicial complexes in a way that preserves their topological invariants, thereby enabling more efficient computations while maintaining the signal components of interest. Although this approach has been explored primarily on synthetic data, we believe that the dialogue between mathematicians and applied researchers could provide valuable new perspectives on the problem of finding compact signal representations.

%% file: Main_text/Sections/5_Conclusions.tex
\section{Conclusions and Outlook}
\label{sec:conclusions}
\noindent \gls{tsp} is an emerging framework for processing signals defined on higher-order, irregular domains in a powerful and elegant way, rooted in principles from algebraic topology and differential geometry. In this work, we have focused on the foundations of \gls{tsp} for signals living on simplicial complexes, based on the Hodge Laplacian operator. We have aimed to present the topic in an accessible manner, highlighting its potential for applications while addressing key questions regarding which types of research questions and data can most benefit from this perspective. To this end, we have introduced the fundamental notions and tools required for such an analysis and illustrated their use through concrete examples. Nevertheless, we have only scratched the surface of a rapidly expanding and still largely unexplored field, which we expect to grow significantly in the coming years. We believe that further progress in this direction will benefit from cross-disciplinary collaborations. In this spirit, we hope that this work may encourage researchers from different communities not to be discouraged by the inherent complexity of the framework, but rather to consider how these ideas might be integrated into their own data analysis pipelines and to foster collaborations across fields.

%% file: Main_text/Acknowledgements.tex
\section{Acknowledgments}
We thank Alexandre Cionca for providing some of the brain icons used in Fig.~\ref{fig:pipeline} and for helpful suggestions on its layout.